\documentclass[preprint2]{aastex}

\slugcomment{To appear in the Astronomical Journal, April 2003 issue}

\shorttitle{{Progenitors of Dwarf Spheroidals}}
\shortauthors{Grebel, Gallagher, \& Harbeck}

\begin{document}

\title{The Progenitors of Dwarf Spheroidal Galaxies}

\author{Eva K. Grebel}
\affil{Max-Planck Institute for Astronomy, K\"onigstuhl 17, D-69117
    Heidelberg, Germany}
\email{grebel@mpia.de}

\author{John S. Gallagher, III}
\affil{University of Wisconsin, Department of Astronomy, 55343 Sterling,
475 N.\ Charter St., Madison, WI 53706-1582}
\email{jsg@astro.wisc.edu}

\author{Daniel Harbeck}
\affil{Max-Planck Institute for Astronomy, K\"onigstuhl 17, D-69117
    Heidelberg, Germany}
\email{dharbeck@mpia.de}

\begin{abstract}
The gas-deficient dwarf spheroidal (dSph) galaxies present an 
evolutionary puzzle that we explore in 40 early-type and late-type dwarfs 
in the Local Group and nearby field.  Although dSphs 
experienced star formation over extended time spans in their 
youths, today all but one are completely free of detectable interstellar 
material, even in the Fornax dSph, where stars formed in the last 
100~Myr. 
Combining photometric and spectroscopic stellar metallicity estimates
for red giant branches with high sensitivity H\,{\sc i} 21-cm line data
from the literature,
we show that the well-established offset in luminosity-metallicity 
relationships for dSphs and dwarf irregular (dIrr) galaxies
exists also when confining the comparison to their old stellar 
populations: 
dSphs have higher mean stellar metallicities for a fixed optical 
luminosity.  Evidently star formation 
in younger dSphs was more vigorous than in the youthful dIrrs,
leading to more efficient enrichment. 
Dwarf galaxies, whose locus in the luminosity-metallicity diagram
is consistent with that of dSphs, even when baryonic luminosities
are considered, are 
the ``transition-type dwarfs'' Phoenix, 
DDO~210, LGS3, Antlia, and KKR\,25. These dwarfs have mixed dIrr/dSph 
morphologies, low stellar masses, low angular momentum,
and H\,{\sc i} contents of at most a few $10^6$ M$_{\odot}$.
Unlike dIrrs, many transition-type dwarfs would closely resemble 
dSphs if their gas were removed, as required to become a dSph; they  
are likely dSph progenitors.   As gas removal is the key factor
for such a transition, we consider the empirical evidence in favor
and against various gas removal processes. 
We suggest that internal gas removal mechanisms are inadequate and favor ram 
pressure stripping to clean the bulk of interstellar matter from galaxies to 
make dSphs.  A combination of initial conditions and environment seems
to support the formation of dSphs:
Nearby dSphs appear to form from small galaxies 
with active early star formation, whose evolution halts due to 
externally induced gas loss. Transition-type dwarfs then 
are dSphs that kept their interstellar medium, and therefore 
should replace dSphs in isolated locations where stripping is 
ineffective.  
\end{abstract}


\section{Introduction}\label{Sect_Intro}

Dwarf spheroidal (dSph) galaxies are the least luminous, least massive
galaxies known. They are defined morphologically in terms of 
their low surface brightnesses, elliptical shapes, and 
relatively smooth distributions of stars.  In galaxy groups,
the dSph galaxies most frequently are found within $\sim 300$
kpc of more massive galaxies, while dSphs exist in large 
numbers as independent members of galaxy clusters.  They have
total masses as low as 
$\sim 10^7$ M$_{\odot}$, $M_V \ga -14$ mag, $\mu_V \ga 22$ mag 
arcsec$^{-2}$, and diameters (from tidal radii) of $\sim 2$ to 
$\sim 6$~kpc.  Their projected radial light profiles are
shallow and can be fit by a range of different profile types,
including exponentials.  Dwarf spheroidal galaxies
are usually devoid of gas and dominated by old and intermediate-age
stars, although some have experienced 
small amounts of recent star formation. Even though dSphs 
can appear to be quite flattened, 
the Galactic dSph satellites are known not to be  
rotationally supported, a trend 
that extends to many, but not all (e.g., Pedraz et al. 2002), 
members of the structurally similar, more massive dwarf elliptical 
(dE) family.  Whether lack of rotational support is a general
characteristic of dSphs is not yet known.
For reviews of dSph properties see, e.g., Gallagher \& Wyse (1994),
Grebel (1997, 1999, 2000), Da Costa (1998), Mateo (1998), and
van den Bergh (1999a, 2000).  Properties of dEs are reviewed by Ferguson 
\& Binggeli (1994).

The dSphs are fascinating galaxies. They combine 
low densities of luminous matter, predilections for locations in dense 
cosmic environments, in some cases high apparent total 
mass densities, and recently recognized complex star formation 
histories. Three general classes of models have been suggested for the  
origins of dSph galaxies: tidal interactions that transform 
field disk galaxies into spheroidal systems (e.g., Sofue 1994; 
Mayer et al. 2001a,b), processes associated with the birth 
of small systems, especially in the cold dark matter 
(CDM) model (Dekel \& Silk 1986),
and fragments liberated during collisions 
between larger galaxies  (Gerola, Carnevali, \& Salpeter
1983). These models 
are not necessarily independent of one another; e.g., dwarfs 
may form via the collapse of CDM and then evolve through tidal 
interactions. While this paper does not study the formation or 
early evolution of galaxies that may become dSphs, we will discuss  
whether multiple evolutionary paths might have dSphs as their 
end points. 

Unfortunately, these perspectives on the origins of dSphs face
well-known difficulties.  For example, standard CDM models
overpredict the fraction of dwarfs which should be dSphs;
the issue in this case is why we do not see
even more examples of this class of galaxy
(e.g., Klypin et al.\ 1999;
Moore et al. 1999; Bullock, Kravtsov, \& Weinberg 2000).
Simple gas loss or
passive evolution to gas exhaustion cannot make a rotating
dIrr into a non-rotating dSph (e.g., Binggeli 1986). While
the tidal model allows angular momentum to be lost, 
it offers neither a ready
explanation for the existence of isolated dSphs, 
such as the Tucana or Cetus dwarfs, nor a simple physical basis 
for the dSph mass-metallicity correlation (see Section \ref{Sect_LZ}).
More generally, if passive evolution leads to
gas exhaustion, we wonder why so few
dwarf galaxies exist with a combination of old stars and neutral gas, as
might be expected if star formation stops at some critical gas density
threshold (Phillipps, Edmunds \& Davies 1990).

While dSphs and dIrrs show similar trends
of global properties with luminosity,
the dSphs are more metal-rich than dIrrs at the same
optical luminosity.  This difference 
is observed both for abundances in planetary nebulae and for stellar
metal abundances, which we denote as ``[Fe/H]'' (e.g.,
Skillman \& Bender 1995; Richer, McCall, \& Stasinska 1998; 
Mateo 1998, and references therein), and stands
as an argument against evolution from dIrrs to dSphs.
Low mass galaxies, including dwarf irregulars (dIrrs) 
and dSphs show similar exponential 
radial brightness profiles (Lin \& Faber 1983;
Kormendy 1985), which has been interpreted as an indication that these 
could be different versions of the same basic type of galaxy. 
In this case another mechanism, such as tides, is required to 
produce spheroidal systems from objects that formed as disk galaxies.
Tidal fragment models for dSphs face different types of difficulties,
such as explaining the apparent lack of extratidal stars, the highly
symmetric structure, and the
high dark matter content of some dSph systems,
such as Draco (e.g., Odenkirchen et al.\ 2001; see also Kleyna ey al.\
2002 for arguments based on a kinematic study).  Note that another 
very nearby dSph, Ursa Minor (Table 1) shows considerable indications
of ongoing tidal disruption (e.g., Mart{\'{\i}}nez-Delgado et al.\ 2001),
showing that at least some dSphs do experience this.

Another key aspect of dSphs is how they achieve their present gas-poor
states (e.g.,  Skillman \& Bender 1995).
In order to have formed stars, all dSphs must once have contained significant
amounts of gas.  However, it is unclear what their progenitors were and
how the transformation from gas-rich to gas-free galaxies proceeded.  
One possibility is that the gas-deficient 
nature of dSphs is closely associated with their births, e.g., 
as a result of special 
effects associated with their rapid early collapse, as predicted by 
CDM galaxy formation models (e.g., Chiu, Gnedin, \& Ostriker 
2001). Alternatively, dSphs could result from 
the transformation of low mass galaxies as predicted 
by models invoking tidal distortion and ram pressure stripping of
objects orbiting near giant systems (Mayer et al. 2001a).  

In this paper we explore the implications of mean stellar population 
metal abundances and H\,{\sc i} gas contents of nearby
low-mass dwarf galaxies, although we include more massive dwarfs 
when considering trends in physical properties.  
We pay particular attention to the  
``transition-type dwarf galaxies'', the dIrr/dSphs.
These objects combine properties associated with dSphs (such as low 
luminosities and prominent
old stellar populations) with those of dIrrs (such as H\,{\sc i} gas). 
The dIrr/dSphs therefore may hold clues to the 
nature of relationships between these two galaxy classes.
In particular we explore whether dIrr/dSph systems 
could be hybrid objects that
merely combine features of dIrr and dSph galaxies, or whether they might 
be in an evolutionary transition from dIrr types towards dSphs (or possibly
the reverse; see Silk, Wyse, \& Shields 1987).

In Section \ref{Sect_Data} we describe observational and derived data on
mean stellar chemical abundances and H\,{\sc i} gas masses of 
dwarf galaxies in the Local Group and its surroundings, which form the 
basis of our analysis.  In
Section \ref{Sect_LZ} the luminosity-metallicity relation for gas-rich dIrr, 
transition-type dIrr/dSph, and gas-free dSph galaxies is discussed.  
Properties of individual transition-type galaxies and selected
dSphs are discussed in 
Section \ref{Sect_simil}, which describes their similarities with respect to
lack of rotation and star formation histories.  
Section \ref{Sect_removal} briefly reviews mechanisms 
that can remove gas from
dwarf galaxies, an essential precondition for producing dSphs.
In Section \ref{Sect_Conc} we present our conclusions.

\section{The nearby dwarf galaxy sample}\label{Sect_Data}

For this investigation we
compiled a sample of nearby dwarf galaxies, whose resolved stellar
populations were studied through deep ground-based observations or 
through Hubble Space Telescope (HST) imaging.
We adopted a generous definition for dwarfs, and 
consider galaxies with $M_V \la -18$ mag to be potential 
members of this structural class (Gallagher 1998; Grebel 1999).
We included all known and suspected Local Group members that lie within
a zero-velocity radius of 1.2 Mpc (Courteau \&
van den Bergh 1999) around the barycenter of the Local Group,
which yields 33 dwarf and satellite
galaxies in addition to the three Local Group spirals.  
The four possible members of the small, nearby Sextans-Antlia
group (van den Bergh 1999b) were added as well.  If Sextans-Antlia
is indeed a physical group, then it is extremely poor,
consisting only of dwarf galaxies. 
Furthermore, we added
three isolated dwarfs in the nearby field.  In total, our sample of 
dwarf galaxies in and around the Local Group 
contains 40 objects, and includes the Magellanic Clouds, 
dIrrs, transition-class dIrr/dSphs, dSphs, and dEs.

Table 1 lists some optical and H\,{\sc i} properties of this sample.  
Horizontal lines separate spatially defined galaxy
subgroups:  The first set of galaxies are possible or likely companions
of the Milky Way, the second companions of M31, the third more distant
possible Local Group members, the fourth possible members of the 
Sextans-Antlia group, and the fifth are 
seemingly isolated galaxies in the nearby field.  
Column 1 gives the dwarf galaxy names,
column 2 their types (see table comments for details), columns 3 and 4
their J2000 coordinates, column 5 heliocentric distances based largely
on the tip of the red giant branch, column 6 the distance to the nearest
``parent'' galaxy (omitted for Milky Way satellites) or to the barycenter
(subscript `bary') 
of the Local Group.  Column 7 specifies absolute $V$-band magnitudes,
column 8 central $V$-band surface brightnesses, column 9 mean metallicity
and metallicity spread (not uncertainty), column 10 the H\,{\sc i} mass,
and column 11 the $V$-band luminosity.  Literature 
references are given in the table comments.  

\subsection{Self-consistent metallicities}\label{Sect_metal}

In order to ensure that we compare the chemical abundances of equivalent 
stellar populations in dIrrs, transition-types, and dSphs, 
we concentrate here on the estimated mean 
stellar [Fe/H] values of old populations.  Old stellar populations
have been detected in each dIrr, transition-type, and dSph 
studied to sufficiently 
faint magnitudes.  We neglect nebular $\alpha$-element
abundances measured in 
H\,{\sc ii} regions or planetary nebulae (PNe).  The former
refer to the {\it present-day} gas abundances of young populations.
The latter form from progenitor
stars with masses $<8 M_{\odot}$ and ages $> 10^8$ years, but cannot
be readily associated with any specific age stellar population.  Since
young metal-rich populations will produce more PNe than older,
metal-poor populations, luminous
PNe help to probe the interstellar
medium oxygen abundances at the epoch when star formation 
last occurred once
certain corrections are applied (Richer, McCall, \& Arimoto 1997).
However, PNe have been detected so far in only two dSphs.  Another
problem arises from variations in the iron-peak to $\alpha$-element abundances,
which depend on prior star formation histories modulated by the details
of gas loss mechanisms (e.g., Burkert \& Ruiz-Lapuente 1997; 
Recchi, Matteucci, \& D'Ercole 2001).  
Hence we consider here only the 
mean metallicities of old stellar populations, which can be 
approximately measured in a consistent manner in both
types of galaxies. 

For old stellar populations in dSphs and dIrrs a large metallicity
data set has become available in recent years based on 
spectroscopic and photometric abundance determinations
for red giants.  Spectroscopic 
measurements have been carried out for individual red giants in a number of
nearby dwarfs (e.g., Suntzeff et al.\ 1993; C\^ot\'e, Oke, \&
Cohen 1999; Bonifacio et al.\ 2000; Guhathakurta, Reitzel,
\& Grebel 2000; Shetrone, C\^ot\'e, \& Sargent 2001; Tolstoy et al.\
2001; Shetrone et al.\ 2003; Smecker-Hane, Bosler, \& Stetson 2003;
Smecker-Hane \& McWilliam 2003).  
Photometric metallicity
estimates can be applied over much larger distances, and are largely based
on the empirical calibration by Da Costa \& Armandroff (1990) and 
more recent refinements (e.g., Lee, Freedman, \& Madore 1993; 
Caldwell et al.\ 1998).

The photometric methods use either the
slope of the red giant branch in comparison with globular cluster
fiducials or the mean $(V-I)_0$ color 
at a certain absolute $I$-band magnitude for old stellar 
populations.  The absolute $I$-band magnitude is chosen such that 
contamination by asymptotic giant branch stars is reduced while 
measuring at a luminosity where the mean color is still sensitive to
metallicity.  This results in metallicities
on the Zinn \& West (1984) [Fe/H] scale, in good agreement with spectroscopic
results.   The absolute accuracy of the photometric estimates is approximately
$\pm 0.2$ dex in galaxies where old stars dominate the red giant 
branch, while the relative accuracy is largely determined by 
photometric uncertainties and can be higher.

An important 
caveat of the photometric methods is the problem of the age-metallicity
degeneracy in mixed populations:  Evolved stars belonging to 
intermediate-age populations overlap with the blue part of the red 
giant branch and may make the red giant branch appear more metal-poor
than it really is.  Synthetic color-magnitude-diagram modelling
helps to reduce this degeneracy if all the various age tracers in a 
color-magnitude diagram are properly taken into account (e.g., Holtzman,
Smith, \& Grillmair 2000).  Also, measurements based on red giant
branches in the outer regions of dIrrs, where old
populations dominate (e.g., Minniti \& Zijlstra 1996;
Harbeck et al.\ 2001), reduce contamination effects.
Still, spectroscopy is the only way to unambiguously resolve the
age-metallicity degeneracy for individual stars.

In Table 1 we list [Fe/H] abundance estimates based on the resolved red
giant branches of nearby dwarf galaxies through intrinsically consistent
methods (spectroscopy where available,
and otherwise photometric estimates).  The $\sigma$ values quoted for 
[Fe/H] refer to the estimated abundance spread across the red giant
branch in these galaxies,
not to the uncertainty in the [Fe/H] determination.

\subsection{Baryon content: Stellar luminosities and 
H\,{\sc i} masses }\label{Sect_baryon}

We used optical $V$-band luminosities ($L_V$) for the galaxies in our sample
(Table 1).  In gas-free dSphs all of the identified baryonic mass is in
stars, whereas some of the baryonic matter in other dwarf galaxies is
still in the form of gas.  To put all dwarfs on
a comparable baryon content
scale, we calculated the ``baryonic luminosity'' $L_{V,\,bary}$
as defined by Matthews, van Driel, \& Gallagher  (1998):
$$L_{V,\,bary} = L_V + 1.33\, M_{\rm HI} \cdot \langle M_V/L_V\rangle ^{-1},$$ 
where $M_{\rm HI}$ denotes the H\,{\sc i} mass, the factor 1.33 accounts
for the presence of helium, and $\langle M_V/L_V\rangle$ is
the mean mass-to-luminosity ratio for the galaxy's stellar population in the 
$V$-band.  We assumed a constant stellar population mass to light ratio 
of 2 to derive the equivalent luminosity of the gas. This is a 
conservative estimate for a fading stellar population (e.g., 
Charlot, Worthey, \& Bressan 1996; Schulz et al. 2002).
Thus $L_{V,\,bary}$ predicts the approximate luminosity that a dIrr 
galaxy would have at the point where it would have converted all of its 
gas into stars and had begun to fade.

For dwarf galaxies with significant amounts of interstellar 
matter, we used the H\,{\sc i} fluxes listed in
Mateo (1998) or more recent publications (see table comments) together
with the distances listed in Table 1 to calculate the H\,{\sc i} masses
via the standard relation 
$$M_{\rm HI} = 2.36 \cdot 10^5\, D^2\, \sum{S_v dv M_{\odot}}$$
(e.g., Knapp, Kerr, \& Williams 1978).  Here $D$ is the galaxy distance in Mpc
and $\sum{S_v dv}$ is the total 21-cm line flux in Jy km s$^{-1}$.  
We revisit the dwarf galaxy H\,{\sc i} content issue in Section~4.2.

Sensitive searches for H\,{\sc i}
within the tidal radii of dSphs resulted in upper column density limits of
only a few $10^{17}$ cm$^{-2}$ (Young 1999; 2000).
The only exception is the
Sculptor dSph, where H\,{\sc i} with peak column densities of $2.2 \cdot
10^{19}$ cm$^{-2}$ was detected within the tidal radius
(Carignan et al.\ 1998).   While location and radial velocity
of the H\,{\sc i} support its association with Sculptor,
this gas may be part of a larger,
unassociated complex merely seen in projection (Carignan 1999).

For gas-deficient dwarf galaxies upper limits are listed.  Where
available these upper
limit masses are derived from H\,{\sc i} line flux limits in  
the recent literature (see table comments for
references), and re-calculated assuming that the H\,{\sc i} fills the 
optical extent of these dwarf galaxies within their tidal radius at uniform
gas column density.  This assumption was made because 
the gas in dwarf galaxies
usually is at least as extended as their optically visible bodies.
In dIrrs the H\,{\sc i} may be several times more
extended than the optical galaxy (see Grebel 2002 for a review and
references).

Non-detections of H\,{\sc i} usually occur in low-mass dwarfs that
are within 300 kpc of more massive galaxies.  Their H\,{\sc i} content,
if present at all, may be truncated in its spatial extent due to
ram pressure effects caused by the massive galaxy's gaseous 
corona, or by tidal
effects (see Section \ref{Sect_removal}).  Hence the
use of the tidal radius appears to be a reasonable compromise in these
gas-deficient objects.  We did not
take into account detections of possibly associated H\,{\sc i} {\em well
beyond} the optical radius of some dSphs (Section \ref{Sect_ram}).  
The comment ``undet.'' refers 
to dSphs that were not detected in the northern Leiden-Dwingeloo Survey (Blitz
\& Robishaw 2000) or in the southern HIPASS survey (Huchtmeier, Karachentsev,
\& Karachentseva 2001), and for which no specific H\,{\sc i}
flux limits were published.

\section{The luminosity--metallicity relationship}\label{Sect_LZ}

Dwarf galaxies follow a well-known luminosity--metallicity relation:
The more luminous a dwarf, the higher its mean metallicity (e.g., 
Grebel \& Guhathakurta 1999).  Higher
luminosities usually imply higher masses and deeper gravitational
potential wells, and/or higher luminous matter densities, and/or
high recent star formation rates.  In return, the more luminous galaxies
may undergo stronger 
enrichment and, due to their deeper potential wells, 
also have an improved ability
to retain the metals (e.g., Dekel \& Silk 1986).

\subsection{Separate branches for dIrrs and dSphs}

If dSphs and dIrrs are distinguished only by evolution induced by 
their environments after formation, then we might expect their older 
stellar populations to be similar. The older stars in both systems presumably 
formed under similar conditions, before the dSph and dIrr evolutionary 
paths diverge. Differentiation by ram pressure stripping alone is a simple 
example of this class of model. The dSph systems would evolve like their 
dIrr cousins until gas is removed, but before gas loss, the two classes 
of galaxies would be similar. Comparisons of chemical abundances in 
older stellar populations therefore can provide powerful evolutionary 
tracers.

\begin{figure}
\plotone{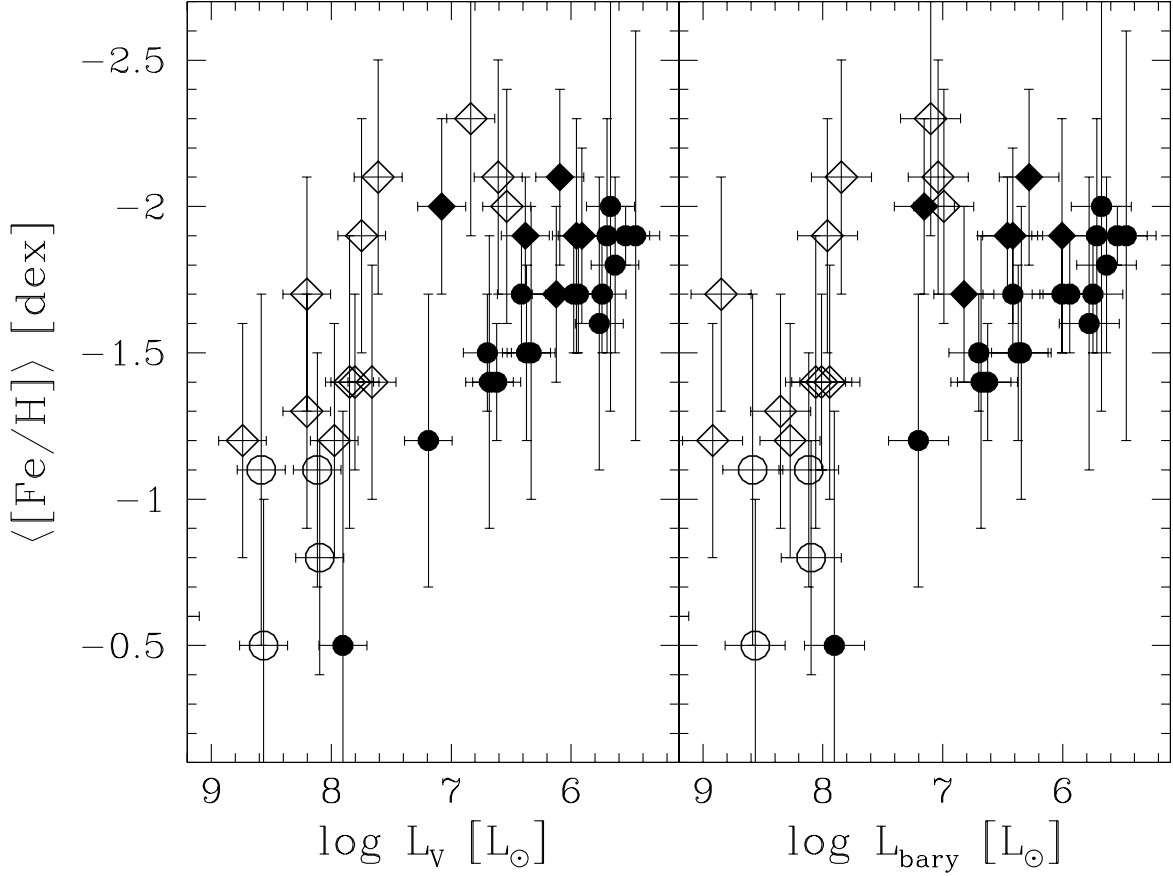}
\caption{$V$-band luminosity (left panel) and baryonic luminosity (right
panel, corrected for baryon contribution of gas not yet turned into stars,
see Section \protect{\ref{Sect_baryon}}
for detail) versus mean metallicity of red giants.
Filled circles stand for dSphs, open circles for dEs, filled diamonds for
dIrr/dSph transition-type galaxies, and open diamonds for dIrrs.
DIrrs are more luminous at equal metallicity than dSphs.  However, several
dIrr/dSph transition-type galaxies coincide with the dSph locus.  These
objects are indistinguishable from dSphs in all their properties except
for gas content. } 
 \label{Fig_LZ}
\end{figure}

A plot of $V$-band luminosity $L_V$ versus
$\langle [Fe/H] \rangle$ (Fig.\ \ref{Fig_LZ}, left panel)
shows a clear trend of increasing luminosity with
increasing mean red giant branch 
metallicity for the dSphs (filled circles), which
appears to extend to the more luminous dEs (open circles).  The dIrrs
(open diamonds) show a similar trend, but are offset from the locus
of the dSphs in that they are more luminous at the same
metallicity (see also Mateo 1998 and references therein).  

In other words, the dIrrs have too low a metallicity for their 
luminosity as compared to dSphs.  Thus dSphs, most of which have been
quiescent over at least the past few Gyr, must have experienced chemical
enrichment faster and more efficiently than dIrrs, which continue to
form stars until the present day.  {\em This is a fundamental difference
between dSphs and dIrrs.}  Plausible progenitors of present-day dSphs
would need to have undergone similarly rapid early evolution.

When plotting the baryonic luminosity against metallicity (Fig.\ \ref{Fig_LZ},
right panel), the locus of
the dSphs remains unchanged while the dIrrs move to
higher luminosities as compared to the dSphs.
Thus if star formation in present-day dIrrs were terminated when all of 
their gas was converted into stars, then these fading dIrrs would be even 
further from the dSph luminosity-metallicity relationship unless substantial 
chemical enrichment occurs in a long-lived stellar population that would 
produce a strong red giant branch.

Figure \ref{Fig_LZ} shows that
present-day dSphs are at least by a factor of 10 fainter in luminosity than
present-day dIrrs at the same stellar metallicity (i.e., for the old
stellar population).  Models presented
by Charlot, Worthey, \& Bressan (1996) show that for a galaxy to 
homologously fade by
this amount --- roughly 2 magnitudes in $M_V$ --- an interval 
of $\ga 15$ Gyr 
after the cessation of star formation is required.  
In other words, present-day dIrrs
would need another Hubble time to turn into dSphs if for some reason
all star formation were to cease now.  For cases where cessation 
of star formation also is associated with loss of stars, e.g., 
in cases of tidal stripping, an even larger decline 
in surface brightness should occur (e.g., Mayer et al.\ 2001a). 
On the other hand, if star formation were to continue
at a low level, as is expected to occur well into the 
future in most dIrrs, due to their modest star formation rates and 
large gas reservoirs (Hunter 1997), the fading time scale would increase.

\begin{figure}
\plotone{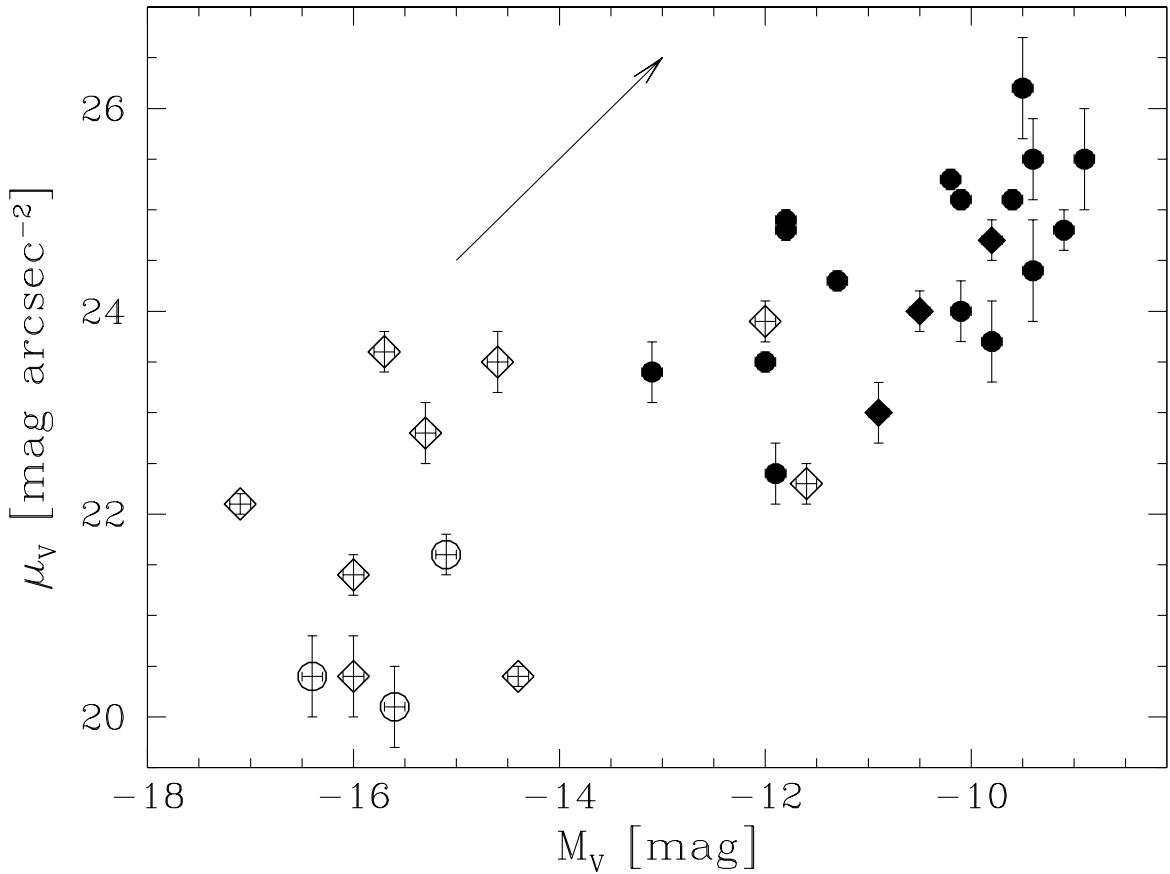}
\caption{Absolute $V$-band magnitude $M_V$ of dwarf galaxies versus their
central $V$-band surface brightness $\mu_V$.  The symbols
for different galaxy types are the same as in Fig.\ 
\protect{\ref{Fig_LZ}}.  Luminous
dIrrs tend to have lower central surface brightnesses than dSphs at the 
same absolute brightness.  The arrow indicates the amount by which
both $M_V$ and $\mu_V$ would be reduced by fading following Hunter \&
Gallagher (1985).  Only low-luminosity dIrrs and transition-type
galaxies would arrive at (or retain) positions consistent with the 
dSph locus in this diagram, while for more luminous dIrrs
fading would result in galaxies with too low a surface brightness
as compared to present-day dSphs. 
} 
 \label{Fig_surf_Mv_FeH}
\end{figure}

At the same absolute magnitude, dIrrs tend to have lower surface
brightnesses than dSphs (Figure \ref{Fig_surf_Mv_FeH}).  If they were
to stop forming stars, fading by one magnitude in absolute brightness
would also imply a decrease in surface brightness by roughly this amount
(Hunter \& Gallagher 1985) as indicated by the arrow in 
Figure \ref{Fig_surf_Mv_FeH}.  For the majority of nearby dIrrs
fading would thus result in galaxies with too low a surface brightness
as compared to present-day dSphs.  A few dIrrs (notably GR\,8 and SagDIG)
as well as the three transition-type dwarfs for which surface brightnesses
have been measured (LGS\,3, DDO\,210, and KKR\,25) would, however, continue to
coincide with the dSph locus in Figure \ref{Fig_surf_Mv_FeH}.

In addition to the issues of fading and rapid early enrichment, many 
dIrrs, and especially the more massive systems, 
would also need to
find a mechanism to lose their angular momentum and their gas if they 
are to become dSphs. We cannot consider most normal present-day 
dIrrs, which are not close satellites of giant 
galaxies, as likely precursors of dSphs over any reasonable  
time scale (see also Binggeli 1986, 1994). 

\subsection{A continuum of transition-type dwarfs?}

Dwarf galaxies classified as
possible dIrr/dSph transition-types in the literature
are marked by filled diamonds in
Fig.\ \ref{Fig_LZ}.
Five of these, namely Phoenix, DDO\,210, LGS\,3, Antlia, and KKR\,25,
lie in
the area of our diagram occupied by the dSphs.  In the $L_{bary}$,
$\langle {\rm [Fe/H]} \rangle$ diagram they outline the luminous edge of the
dSph locus.  Their H\,{\sc i} masses are typically a few times $10^6$
$M_{\odot}$, while dIrrs have H\,{\sc i} masses $\ge 10^7$
$M_{\odot}$.  

The  sixth potential transition-type galaxy, PegDIG,
is found in the region of Fig.\ \ref{Fig_LZ} occupied by the dIrrs.
Two metal-poor, faint dIrrs lie close to the dSph locus
in the $L_V$, $\langle [Fe/H] \rangle$ diagram:
GR\,8 and Leo\,A.  They move closer to the dIrr locus when $L_{bary}$
is plotted instead, although since Leo~A has a dominant moderately young 
stellar population (Tolstoy et al.\ 1998), its stellar [Fe/H] may be 
underestimated (cf.\ Schulte-Ladbeck et al.\ 2002). We also tentatively 
exclude the PegDIG from the 
transition group due to the nature and level of its star-forming activity 
and substantial interstellar medium (see Section \ref{Sect_difftransstage}).

The locations of Phoenix, DDO\,210, LGS\,3, and Antlia on the 
luminosity-metallicity plane indicate that
these galaxies would closely
resemble the classic dSph galaxies if they were to lose their gas.
Indeed, in the $L_V$, $\langle [Fe/H] \rangle$ diagram they would
be essentially indistinguishable from dSphs.  In this sense, we
may consider these
four transition galaxies as present-day progenitors of dSph galaxies, whose
properties (other than gas content and associated recent
star formation) are fully consistent with the
properties of dSphs. However, given the reasonable H\,{\sc i} contents 
and low star formation rates
of these galaxies, they are not rapidly evolving. 

The nearby transition-type galaxies appear to close the
gap between the late-type and early-type dwarf galaxies at the
low-luminosity end, possibly even indicating the existence of a continuum
of faint, low-mass dwarf galaxies between the dIrr and dSph 
morphological classes.

\section{Properties of transition-type dwarfs}\label{Sect_simil}

Are the overall properties of transition-type galaxies consistent with their 
being  progenitors of dSph galaxies?  

\subsection{Absence of rotation}

If the ratio of rotational velocity $v_{\rm rot}$ versus velocity
dispersion $\sigma$ is $\la 1$, random motions rather than rotation
dominate (e.g., Lo, Sargent, \& Young 1993).  This appears to be the
case for all dSphs whose stellar kinematics have been studied to date.

In contrast to more massive dIrrs, most of the low-mass dIrr and 
dIrr/dSph galaxies show little evidence for  
rotational support; e.g.,  GR\,8, Leo\,A, LGS\,3, SagDIG, and 
DDO\,210 do not
display systematic rotation in their  H\,{\sc i} 
gas (Lo et al.\ 1993;
Young \& Lo 1997).  The H\,{\sc i} cloud apparently associated with
Phoenix has a small velocity gradient that may either be due to rotation
($v_{\rm rot}\, \sigma^{-1} = 1.7$) or due
to ejection from the galaxy (St-Germain et al.\ 1999).  PegDIG also
has $v_{\rm rot}\, \sigma^{-1} = 1.7$ (Lo et al.\ 1993).
No such data are available yet for Antlia and KKR\,25.   
These H\,{\sc i} velocity field measurements 
are consistent with either no ordered circular motion or
very low levels of rotation ($v_{\rm rot} \le$ 5~km~s$^{-1}$), 
supporting close kinship between dSphs
and dIrr/dSphs, although obviously observations of {\em stellar} 
kinematics are essential to confirm this point. The need for further 
measurements of stellar velocities to constrain rotation also applies 
to many of the Local Group dSphs (e.g., the M31 dSph companions).

\subsection{Distances and H\,{\sc i} masses}

\begin{figure}
\plotone{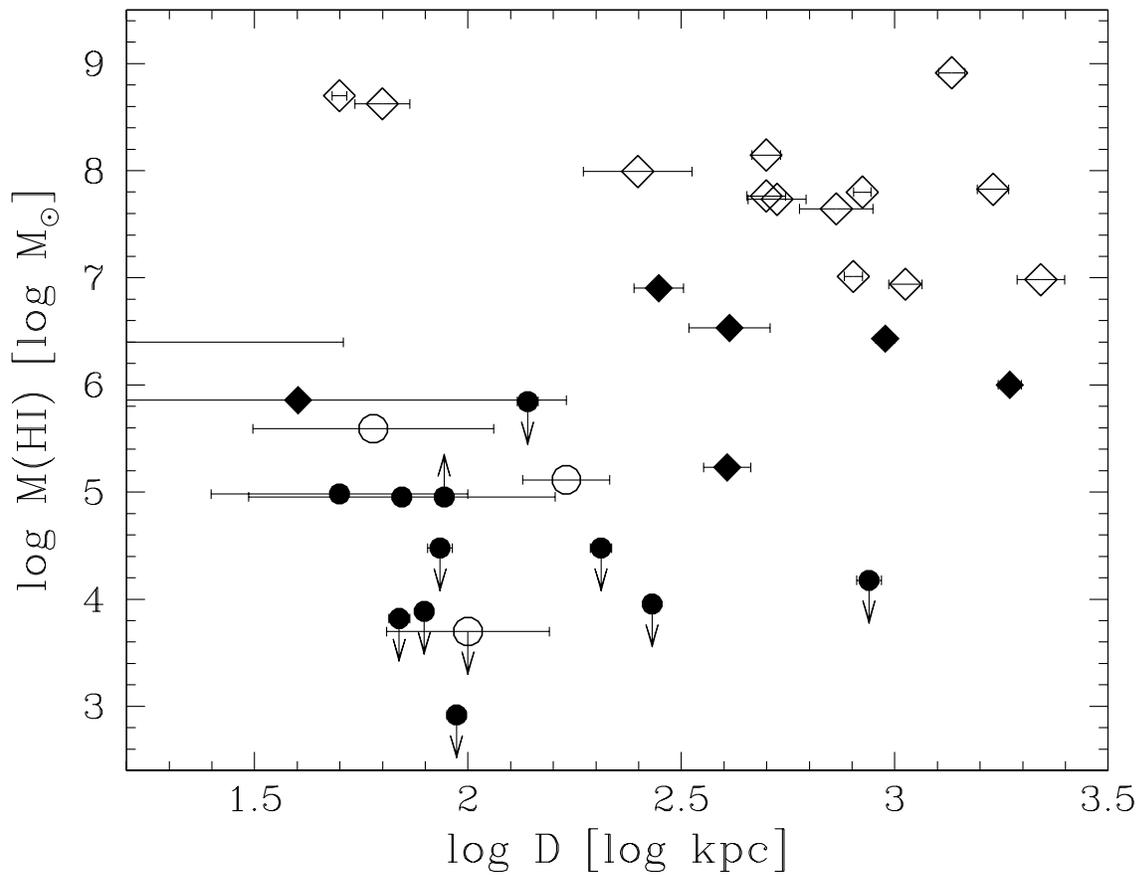}
\caption{Dwarf galaxy
H\,{\sc i} mass versus distance to the nearest massive galaxy. The symbols
are the same as in Fig.\ \protect{\ref{Fig_LZ}}.  Lower or
upper H\,{\sc i} mass limits are indicated by arrows. There is a 
general trend for the H\,{\sc i} masses to increase with increasing distance
from massive galaxies.  DSphs lie typically below $10^5$ M$_{\odot}$ in 
H\,{\sc i} mass limits, while potential transition-type galaxies have
H\,{\sc i} masses of $\sim10^5$ to 10$^7$ M$_{\odot}$. DIrr galaxies
usually exceed  10$^7$ M$_{\odot}$.}
 \label{Fig_Dist_MHI}
\end{figure}

Plotting the H\,{\sc i} masses in Table 1 
versus distance from the closest massive
galaxy (column 5 in Table 1), we see a tendency for H\,{\sc i} masses
to increase with galactocentric distance (Figure \ref{Fig_Dist_MHI}).  
This trend was also noticed
by Blitz \& Robishaw (2000), who confirmed the finding by
Knapp, Kerr, \& Bowers (1978) that for dSphs H\,{\sc i} mass upper limits
fall well below 10$^5$ M$_{\odot}$.  Only fairly large 
galaxies, such as the Magellanic Clouds, IC\,10, and
M33, with H\,{\sc i} masses $\gg 10^7$~
M$_{\odot}$, seem to be able to retain their gas reservoirs 
when closer than $\sim 250$~kpc to giants.
We note that weak lensing measurements and dynamical modelling indicate
typical dark matter halo scales of $260 h^{-1}$ (e.g., McKay et al.\ 2002),
of similar order as the distance range discussed here.
These H\,{\sc i}-rich satellite galaxies are at least two orders of
magnitude more massive than the dSphs.
The one transition-type galaxy within this distance
range of a larger neighbor is Antlia, which is 
believed to be tidally interacting with NGC\,3109 (Section \ref{Sect_tidal}).

Beyond distances of $\sim$250~kpc from the two Local Group giant galaxies, a 
number of dIrrs and transition-type galaxies with substantial H\,{\sc i} mass
fractions are found.  
While the absence of such galaxies at smaller distances seems
consistent with expectations from a distance-dependent
ram pressure/tidal stripping scenario (see also Section \ref{Sect_external}),
we note that dIrrs and dSphs are seen throughout the Local Group,
although very few dSphs are known at distances $\ga 250$~kpc are known.
Environment cannot be the only factor 
responsible for the origin of dSph galaxies (see also Tamura \& 
Hirashita 1999). 

Detections of possibly associated H\,{\sc i} clouds at distances of $\sim
10$ kpc from the optical centers of five dSphs were reported by
Blitz \& Robishaw (2000; cf.\ Oosterloo, Da Costa, \&
Staveley-Smith 1996).  Could these be
former dIrr/dSphs, which were stripped recently?

For Andromeda\,V, the claim of physically associated gas can be excluded
due to a difference between the optical and H\,{\sc i}
radial velocity (Guhathakurta, Reitzel, \& Grebel 2000).  The recent 
high-resolution H\,{\sc i} study by Robishaw, Simon, \& Blitz (2002)
shows that the 
detection reported near Leo\,I is kinematically distinct from this
galaxy.  The authors suggest, however, that this cloud may be tidally
interacting with Leo\,I. 
Matching the H\,{\sc i} cloud seen projected in the vicinity of
Tucana to this galaxy requires a measurement of an
optical velocity, which is still missing.
Oosterloo et al.\  (1996)
consider it more likely that this cloud
is either a high-velocity cloud or part of the Magellanic
stream.  The remaining two H\,{\sc i}
detections have velocities similar to the optical radial velocities of
the adjacent dSphs, And\,III (Guhathakurta et al.\ 2000) and Sextans, which
may indicate a connection with these dSphs, although the
caveat of possible
confusion with intergalactic gas clouds should be kept in mind.

None of the three dSphs with possibly associated  
H\,{\sc i} (And\,III, Sextans, Tucana) contains
either prominent intermediate-age ($<10$ Gyr) populations or young stars.
The distances of these three dSphs from the closest spiral galaxy are $\sim 70$
kpc (And\,III), $\sim 86$ kpc (Sextans), and $\sim 870$ kpc (Tucana);
distance is not the common factor.
Blitz \& Robishaw (2000) suggest that
ram pressure stripping is the most probable cause for the removal
of the gas, if the nearby H\,{\sc i} clouds were indeed
once part of the dSphs.
However, at Tucana's location, near the edge of the Local
Group, there is no obvious agent for ram pressure stripping other than
a hypothetical, highly inhomogeneous intergalactic medium (Section
\ref{Sect_clumpy}).

\subsection{Gas and star formation histories}

The available data suggest that dSphs and transition dwarfs share similar star
formation histories.  Both contain 
substantial old stellar populations with ages $> 10$ Gyr.  
A number of dSphs and all transition-type
galaxies also contain intermediate-age populations ($< 10$ Gyr).
The present-day star formation activity in transition-type dwarfs, 
when present, occurs at low levels. Typical
values are a few times $10^{-3} M_{\odot}$ yr$^{-1}$ (Mateo 1998), which 
are sufficient to produce the observed stellar populations if continued 
over a cosmic time span. However, some transition dwarfs could exhaust the 
observed gas supplies in only a few Gyr, suggesting that, unlike the 
dIrrs, the transition-type galaxies may be approaching the end of their 
lives as actively star-forming galaxies.

In all cases, the mean metallicities are low $\langle [Fe/H] \rangle < 1$ dex).
Metallicity spreads are observed among the red giant
branch stars, indicating efficient enrichment processes within the old
population and presumably similar modes of star formation.
{\it The only significant difference between integrated properties of the 
classic dSphs and transition-type dwarfs appears to be
the absence or presence of gas and the  associated recent 
star formation rate.} We therefore may ask whether it is the presence 
or absence of gas that would be the norm in a small galaxy. Theoretical 
models support the former option; most dwarf galaxy models do not evolve to 
gas-free states (e.g., Mac Low \& Ferrara 1999; Andersen \& Burkert 2000; 
Carraro et al.\ 2001). In the remainder of this section we consider 
the observational case regarding this issue.

\subsubsection{Galaxies in different stages of transition?}\label{Sect_difftransstage}

A comparison of four well-studied low-mass dwarf galaxies provides 
a way to more closely examine relationships between dSph and transition-type 
galaxies:

The location of neutral gas in PegDIG is well correlated with the optical
structure of the galaxy, and  H\,{\sc i} concentrates
near areas of recent
star formation (Lo et al.\ 1993).  Star formation in this dwarf galaxy
at a distance of $\sim 410$ kpc from M31
has continued at a roughly constant rate 
until today (Gallagher et al.\ 1998), and its H\,{\sc i}
mass is an order of magnitude higher than in Phoenix and LGS\,3.
Despite its dIrr/dSph morphology, this galaxy does not appear to
be a likely progenitor of a dSph since its properties are consistent
with those of low-mass dIrrs.

The transition-type galaxy LGS\,3 is at
a distance of $\sim 280$ kpc from both M31 and M33.  
It is unclear whether it is a companion
of one of these spirals, but the larger mass of M31 implies
that LGS\,3 may belong to M31 (Miller et al.\ 2001).
In optical images LGS\,3 resembles PegDIG.  However, while
the H\,{\sc i} in LGS\,3 is roughly symmetrically distributed around 
its optical center, its H\,{\sc i} mass
is only one tenth of the stellar mass (Young \& Lo 1997).
LGS\,3 formed the bulk of its stars very early on, and subsequent star
formation proceeded at a continuous, but decreasing rate (Miller et al.\ 2001).

Phoenix is a transition-type galaxy at a
distance of $\sim 400$ kpc from the Milky Way.
It experienced fairly continuous star formation until
approximately 100--200 Myr ago (Holtzman, Smith, \& Grillmair 2000),
but, like LGS\,3, appears to have
formed most of its stars long ago.   It 
is associated with an off-centered $\sim 10^5$ M$_{\odot}$
H\,{\sc i} cloud (St-Germain et al.\ 1999; Gallart et al.\ 2001).
While the regular structure of this H\,{\sc i} cloud argues against
expulsion due to supernova explosions, Gallart et al.\ (2001) suggest
that ongoing ram pressure stripping may be removing the gas from Phoenix.
More recent work (Irwin \& Tolstoy 2002), however, shows that 
the velocity of the cloud and the optical velocity of the galaxy 
are in close agreement, making this scenario seem less likely.

The Fornax dSph (distance from the Milky Way $\sim 140$ kpc) contains a
sizeable old stellar population, as evidenced 
through the detection of numerous
RR Lyrae variables (Stetson, Hesser, \& Smecker-Hane
1998; Bersier \& Wood 2002).  It is one of only two dSphs in
the Local Group with ancient globular clusters (e.g., Buonanno et al.\
1998), but the majority
of its stars appear to be of intermediate age.  Fornax experienced a 
decreasing level of star formation until stellar births ceased 
100--200 Myr ago, making it
the dSph with the most recent star formation (Stetson et al.\ 1998;
Grebel \& Stetson 1999).  The nearby Fornax system
is the largest, most massive,
and most luminous galaxy of the four dwarfs discussed in this 
subsection, and experienced the strongest chemical enrichment.
That it contains a young population is unusual for a dSph 
and indicates that
the removal of its star-forming interstellar medium 
must have occurred recently.  The recent proper motion estimate
by Piatek et al.\ (2002) indicates that Fornax is currently close
to perigalacticon and may be bound to the Local Group rather than
to the Milky Way (see Section \ref{Sect_ram}).

Fornax,
Phoenix, and LGS\,3 closely resemble each other in several ways:  All
experienced star formation until 100--200 Myr ago, all have prominent
intermediate-age populations as evidenced by, e.g., their red clumps and
subgiant branches,
and they all have substantial old populations traced by well-populated
horizontal branches.  None of the three dwarfs shows evidence
for rotation, and the H\,{\sc i} mass in Phoenix and LGS\,3 amounts only
to a few $10^5$ M$_{\odot}$.  Phoenix and LGS\,3 could easily
pass as dSphs if they did not contain measurable amounts of gas, and 
Phoenix may soon reach a dSph-like gas deficiency, if its gas is indeed in
the process of being removed.
Do these three dwarfs then represent galaxies in
different phases of transition into purely gas-free dSph systems?

\subsubsection{The transition-type dwarf candidates at larger distances}

Less detailed information is available for the other transition-type
galaxies discussed in Section \ref{Sect_LZ}, since observations of the
depth needed to constrain star formation histories have yet to be made.

DDO\,210 
experienced star formation until $\sim 100$ Myr ago (Lee et al.\ 1999).
While it does contain neutral gas, its H\,{\sc i} is asymmetrically  
distributed with
respect to the optical center of the galaxy; 
the  H\,{\sc i} is clumpy, and shows a larger
angular extent to the north (Lo et al.\ 1993).  The H\,{\sc i}
comprises one half or one third of the estimated total mass
of DDO~210.

KKR\,25 is a very isolated transition-type galaxy at a distance of 
$\sim 1.9$ Mpc
in the direction of the Hercules--Aquila Void (Karachentsev et al.\
2001).  It resembles the other transition-type galaxies in stellar
content. Its small H\,{\sc i} mass is roughly half of its dynamical
mass (leaving remarkably little room for a dense dark matter 
halo).  Its star formation history is poorly known owing
to its distance. KKR\,25, however, is an important object that has 
transition-type characteristics, but is very far from giant galaxies; 
its transition status presumably is intrinsic to the galaxy.

\subsubsection{Isolated dSphs}

While the trends in morphological properties,
gas content and in star formation history with 
distance appear to support
a scenario in which transitions between low-mass galaxies are aided
by environmental effects (see also van den Bergh 1994; Grebel 1997), 
this scenario cannot explain the existence of the gas-deficient dSphs
Tucana and Cetus at large distances from the 
Local Group spirals.  Both these dwarfs 
are dominated by old stellar populations and do not contain prominent 
intermediate-age or young populations.  The star
formation histories of these two dSphs indicate that their early evolution 
was governed by extended episodes of star formation that led to 
considerable metallicity spreads.  
These two dSphs even show radial
population gradients in their old 
stellar populations, while Tucana may have a bimodal 
metallicity distribution that could be indicative of 
multiple starbursts (Harbeck et al.\ 2001;
Sarajedini et al.\ 2002).  

In the luminosity-metallicity diagram, Tucana and Cetus 
are indistinguishable from other dSphs.
Their existence demonstrates that present-day environment cannot be 
the only factor responsible for the existence of dSph galaxies.

\section{Gas removal mechanisms}\label{Sect_removal}

Observations and theory appear to agree in predicting that most 
dwarf galaxies should be able to retain a reasonable fraction of 
their primordial gas supplies.  
If evolution from transition-type dwarfs to dSphs is to occur, 
then effective mechanisms must exist to remove interstellar gas.  
In the following
we briefly review several possible gas removal mechanisms.

\subsection{Internal mechanisms}\label{Sect_internal}

\subsubsection{Gas expulsion through star formation}

 While star formation may certainly lead to
galactic winds
and significant metal losses from dwarf galaxies, numerical
models indicate that star formation is unlikely to be
a generally valid agent for the {\em complete} removal of gas.
Stellar-powered blow-away of the entire interstellar medium 
is predicted to occur
only in galaxies with total masses below $5\cdot 10^6$ M$_{\odot}$
(Burkert \& Ruiz-Lapuente 1997; Hirashita 1999; Mac Low \& Ferrara 1999; 
Andersen \& Burkert 2000; Ferrara \& Tolstoy 2000).
However, no galaxies with such low total masses are known.  The inferred
total masses of dSphs comfortably exceed this limit, assuming that
they contain dark matter (Mateo 1998).
The assumption of internal virialisation leads to essentially the
same estimated dynamical mass for all
Galactic dSphs ($10^7$ M$_{\odot}$; Gallagher \& Wyse 1994). 

Furthermore, neither the transition-type dwarfs nor the dSphs show compelling 
evidence for the major starbursts needed to produce 
large-scale gas blow-outs in their star formation histories (to the extent
that such events would be resolvable).  In particular, there is no evidence
of dSphs whose star formation terminated after a major starburst.
With the exception of the
clearly episodic star formation history of the Carina dSph (Smecker-Hane 
1997; Hurley-Keller, Mateo, \& Nemec 1998), the 
star formation histories of dSphs
are consistent with fairly continuous star formation 
whose rate subsided at recent times (see Grebel 1999;
2000 for reviews).  

This is consistent with the existence of dwarf galaxies with apparently 
similar masses to those of the dSphs, but with substantial H\,{\sc i} 
gas reservoirs (e.g., LGS3, DDO210, and KKR25; Table 1).
Moreover, the large metallicity spreads observed in a number of
dSphs that are dominated by ancient star formation episodes
(Shetrone et al.\ 2001, 2003; Ikuta \& Arimoto 2002;
Carigi, Hernandez \& Gilmore 2002)
as well as radial gradients in old populations (Harbeck et al.\ 2001)
indicate that dSphs were not only able to hold on to a significant
amount of the metals that they produced, but also experienced extended
ancient star formation episodes.  Recent detailed analyses of
the age-metallicity evolution favor leaky-box models (e.g., Smecker-Hane \&
McWilliam 2003) or even closed-box models (e.g., Tolstoy
et al.\ 2003).

\subsubsection{Gas exhaustion by star formation}

Davies \& Phillips (1988) suggest that
the gas in dSphs was consumed completely by star formation.
However,
star formation efficiencies are usually fairly low, typically in the 
range from 1\% to 9\% per event in Galactic giant molecular clouds
(Carpenter 2000).  Even though upper limits for the efficiency of
low-mass star formation may reach 30\% to 50\% when the effects of massive
stars are neglected (Matzner \& McKee 2000), one expects to find that
much of the gas remains after stars form. Thus we observe the common pattern 
where the mass of interstellar gas exceeds that of recently formed 
stars.

Furthermore, in order for star formation to occur, the gas density
probably must exceed some critical value to produce gravitationally
bound gas clouds where stars can condense.  Galaxies with low gas
pressure are dominated by warm
neutral interstellar matter (ISM; with temperatures $\la 10^4$ K).
The gas can cool down and form molecular clouds only in regions with
localized high pressure.  In absence of shear
and density waves, turbulence may locally lead to long-lived clumps with
higher density (Elmegreen \& Hunter 2000).  In this 
model the condensation of stars occurs in 
local density peaks and ceases once 
the gas density becomes lower than some critical threshold for 
gravitational collapse (van Zee et al.\ 1997). This results in
low global efficiency, scattered star formation that cannot 
consume all of the available star-forming gas. 

A dwarf galaxy left on its own therefore should evolve 
into an object with little or no active star formation, even 
though significant amounts   
residual gas would remain (e.g., see models of Carraro et al.\ 2001).  

\subsubsection{Gas replenishment}\label{Sect_replen}

The absence of detectable amounts of 
interstellar gas in dSphs is surprising since 
the slow replenishment of the
ISM naturally will occur due to mass ejected  
from stars in the course of their 
evolution.  Stellar population models with normal initial mass 
mass functions and low metallicities show that stellar mass loss rates 
decline as the population ages, scaling approximately as $t^{-1}$ 
at late times (e.g., Jungwiert, Combes, \& Paulou\v{s} 2001). The Schulz et al.
(2002) simple stellar population models give a total mass return of 
$\approx$ 5\% of the present day stellar mass in a system that is 
passively aging from 4--12~Gyr after its formation. Limits 
on the H\,{\sc i} masses in dSphs typically lie in the range of 
0.1\% to 1\% of the current stellar mass (see Table 1). Detectable amounts 
of H\,{\sc i} then should build up in dSphs in at most a few Gyr 
if stellar mass loss is retained and becomes neutral (Young 2000). 
Yet this is not observed.  

Hence we face two substantial questions: How and why did 
dSph progenitors lose their initial supplies of gas?  What prevented 
their refilling with gas lost by stars to the point where they could be 
detected in the recent, sensitive 21~cm H\,{\sc i} line surveys (see also
Sections \ref{Sect_external}, \ref{Sect_ion})?

\subsection{External mechanisms}\label{Sect_external}

\subsubsection{Gas removal through ram pressure stripping?}\label{Sect_ram}

Ram pressure stripping long has been suspected to play a major role 
in producing gas-deficient dSph galaxies (e.g., Eskridge 1988, van den Bergh 
1994). 
In order for ram pressure stripping to occur, the following condition
has to be met (e.g., Gunn \& Gott 1972):
$$P_{\rm ram} \sim \rho_{\rm IGM}\, v^2 > \frac{\sigma^2 \rho_{\rm gas}}{3}.$$ 
Here
$P_{\rm ram}$ denotes the ram pressure, $\rho_{\rm IGM}$ the gas density
in the intragalactic medium (IGM), $v$ the velocity of the galaxy through
the IGM, $\sigma$ the galaxy's velocity dispersion, and $\rho_{\rm gas}$ 
its gas density. Use of this simple estimate for ram pressure stripping 
is supported by numerical models (e.g., Mori \& Burkert 2001; Quilis \& 
Moore 2001).

Murali (2000) estimates a mean gaseous halo density of 
$n_H < 10^{-5}$ cm$^{-3}$ at 50 kpc from the Milky Way, 
which implies that at larger distances  the Local Group IGM
density should be even lower.  Dwarf galaxy radial
velocities set a lower limit for their space velocities, which likely
do not exceed a few hundred km s$^{-1}$ in order for them to 
remain bound to the Milky Way or M31.  Ram pressure stripping 
is favored by proximity to giant galaxies, where halo gas densities 
may already exceed $10^{-4}$ cm$^{-3}$ at distances of 
several tens of kpc (Stanimirovi\'c et al. 
2002), and where  
relative orbital velocities will be highest.  

Is the dSph Fornax (Section 
\ref{Sect_difftransstage}) then a galaxy that could have
experienced ram pressure stripping recently?  Piatek
et al.\ (2002) suggest that Fornax 
is close to perigalacticon, that it has a large
tangential velocity ($310\pm80$ km s$^{-1}$), and that may not be bound
to the Milky Way.  While the uncertainties of the proper motion 
measurement are still large, we use the resulting space velocity of
Fornax, its stellar velocity dispersion of $\sim 10$ km s$^{-1}$ (Mateo
1998), and $\rho_{\rm IGM} < 10^{-5}$ cm$^{-3}$ to estimate the 
internal average gas density below which ram pressure stripping could
have occurred:  We find $\rho_{\rm gas} \la 10^{-2}$ cm$^{-3}$,
corresponding to an average column density of $\la 10^{19}$ cm$^{-2}$. 
We note that the IGM gas density is a conservative upper limit
considering that Fornax is at $\sim 3$ times the distance of the 
Magellanic Stream.  Moreover, Fornax'  negative Galactocentric
radial velocity indicates that is still approaching perigalacticon
and has not yet passed through denser parts of the Galactic halo.
Hence it appears questionable whether ram pressure stripping through
a homogeneous gaseous medium could have removed Fornax' gas so
efficiently.

Internal gas densities differ from
dwarf galaxy to dwarf galaxy.  In Phoenix, the 
{\it peak} H\,{\sc i} column density is a few times
$\sim 10^{19}$ cm$^{-2}$ or
$\rho_{gas} \sim 3\cdot 10^{-3}$ cm$^{-3}$.  In low-mass dIrrs and
dSphs the internal velocity dispersions are
$\sigma \la 10$ km s$^{-1}$.  We assume furthermore that the IGM is
homogeneous.  Our order-of-magnitude estimate using the above formula
shows that for interstellar gas surface densities of 
$\sim 10^{19}$ cm$^{-2}$ the ram pressure is roughly balanced within 
a typical dwarf galaxy located in a galactic halo with 
$n_H \sim 10^{-5}$ cm$^{-3}$.  For galaxies with lower
interstellar gas densities, ram pressure stripping in galactic 
halos becomes increasingly effective.  
Since dwarf
galaxies usually have heterogeneous interstellar media with varying densities,
ram pressure may strip gas from their 
least dense regions when they encounter a relatively dense gas in a 
halo or the IGM (see Bureau 
\& Carignan 2002). 

The present-day mean density of the IGM beyond galaxy halos in the Local Group 
is thought to be $n_H \le 10^{-5}$ cm$^{-3}$, which is too low to 
remove H\,{\sc i} even from dIrr/dSph dwarfs (see Quilis \& Moore 2001),
unless peculiar velocities are very high.   
This problem is more severe for dIrr galaxies. For example, in Leo\,A 
with its  
peak H\,{\sc i} column densities of $> 10^{20}$ cm$^{-2}$, 
ram pressure due to any uniform Local Group IGM will have
little effect on the gas.  So the question of how transition 
dwarfs (let alone dSphs) became as gas-poor as is observed today remains 
unanswered. 

A partial solution to this problem possibly lies in the evolution
of the Local Group IGM. In the past the IGM may have been denser and perhaps 
also clumpier.  During the formation starbursts 
of giant galaxies, galactic winds would have been
strong, giant galaxies probably had larger `spheres of influence'
within which the IGM would be highly perturbed,  and 
dwarf galaxies could be stripped (Hirashita, Kamaya, \& Mineshige 
1997; Scannapieco, Ferrara, \& Broadhurst 2000).

An obvious shortcoming of models that rely on proximity to giant galaxies
for gas removal from dwarf systems is the existence of the Cetus and
Tucana dSphs far from the Milky Way and M31. One possibility is that
both of these dwarfs once were near one of the Local 
Group giants, where they were stripped of their gas, but 
subsequently escaped towards their
present locations. However, the time for them to reach their present
day locations from the inner Local Group would be 6 to 10~Gyr for a
space velocity of 100~km~s$^{-1}$, and so gas refill from stellar 
mass loss is an issue (Section \ref{Sect_replen}).

This difficulty might be overcome if following stripping of an
unbound dwarf galaxy in
a close passage near a giant, the Local
Group IGM could remove stellar mass loss as it is injected.
A rough estimate can be made by assuming that stellar mass loss
fills an initially gas-free dwarf in a sound crossing time of 100 to 
300~Myr. In this case the gas density within a dwarf like
Cetus or Tucana is $\lesssim 10^{-5}$~cm$^{-3}$ with an ISM mass
corresponding to about 1000~M$_{\sun}$. Even a very low-density
IGM with $\rho_{IGM} \sim 10^{-7}$ should be able to produce
ram pressure stripping under these conditions in the few hundred
Myr time scale.

\subsubsection{Tidal stripping in progress?}\label{Sect_tidal}

Apart from ram pressure stripping, tidal stripping may be a valid
mechanism to rid low-mass galaxies of their gas {\it if} their orbits
bring them near enough to massive galaxies.
The dIrr/dSph galaxy Antlia is a close ($\ge 40$ kpc)
companion of the more massive, gas-rich dIrr
NGC\,3109 (van den Bergh 1999b).  Antlia
appears to have had a close encounter with
NGC\,3109 about 1 Gyr ago,
which caused a warp in the disk of NGC\,3109 (Barnes \& de Blok 2001).
It now has a low H\,{\sc i} mass of $\sim 7\cdot 10^5$
M$_{\odot}$.
How much gas was lost from Antlia during the encounter is unknown, but
Antlia may be a good candidate for tidal gas stripping.

However, tidal stripping does not seem to be a universally valid agent for the
removal of gas, unless the more distant dSph and transition-type
systems are on extremely
eccentric orbits. And even in this case replenishment of their interstellar
medium by mass lost from evolving stars will likely present a problem
due to the long time interval between any hypothetical perigalacticon
passages.  An example of a gas-devoid dSph that shows no signature of
tidal stripping is the strongly dark-matter-dominated, nearby Milky Way
satellite Draco (Odenkirchen et al.\ 2001), one of the closest Galactic dSphs.

\subsubsection{Stripping by the Local Group IGM?}\label{Sect_clumpy}

An alternative possibility for explaining the free-flying 
Local Group dSphs could be an inhomogeneous IGM . If the IGM is 
heterogeneous and contains regions where $n_H > 10^{-5}$
cm$^{-3}$, then encounters with such clumps could remove H\,{\sc i} gas in 
proto-dSphs that are far from giant galaxies.
The details will depend on the geometry of the encounter.
For instance, it has been suggested that the supergiant H\,{\sc i} shell 
structure observed in the dIrr NGC 6822 may have been caused by interaction
with a massive H\,{\sc i} cloud (de Blok \& Walter 2000).   
This idea was also suggested as a model for the off-centered
H\,{\sc i} in the Phoenix dSph/dIrr by Gallart et al.\ (2001) and 
remains a viable option even after the Irwin \& Tolstoy (2002) 
revision of the galaxy's radial velocity. The necessary
structure in the Local Group IGM could, for example, result from
disturbances associated with the passage of galaxies through the IGM
(see Silk, Wyse, \& Shields 1987).  Another possibility is that initial
rapid stripping of gas from some dwarfs occurred at relatively early
times, when the IGM may have been denser and more highly clumped.

Is there evidence for a heterogeneous IGM in the Local Group?
High-velocity clouds 
appear to be mostly local objects in close proximity (i.e., within
several kpc) to massive
galaxies such as the Milky Way (Wakker 2001), which  makes them
unlikely agents for stripping of the more distant dSph galaxies. 
Compact high-velocity clouds may be at larger distances (e.g., Braun \&
Burton 2001), but recent HIPASS results (Putman et
al.\ 2002) imply that they are largely part
of local features, such as the Magellanic stream.  As such they may
be capable of affecting nearby dSphs.  Previously reported H\,{\sc i}
clouds without optical counterparts in galaxy clusters could not be
confirmed (van Driel et al.\ 2002, and references therein).
Searches of nearby
groups of galaxies for H\,{\sc i} clouds down to masses of $3 \cdot 10^6$
M$_{\odot}$ yielded non-detections (de Blok et al.\ 2002), indicating
that the present-day 
IGM is not usually clumpy on these mass and density 
scales.  The lowest column density
limits reached so far are $\sim 10^{18}$ cm$^{-2}$ (Zwaan 2001;
corresponding to upper bound density limits of 
a few times $10^{-4}$ cm$^{-3}$ if sizes of 1 kpc are
assumed), which appears to render a highly  
clumpy IGM at the present epoch less likely.  Less massive
clumps or more wide-spread IGM inhomogeneities at earlier epochs, however, 
are not ruled out.

At low densities the IGM will be photoionized by the UV background radiation 
in combination with any local sources. Gas in this regime will not show up 
in H\,{\sc i} surveys. However, UV and X-ray line studies can find such 
material through its line absorption if it contains metals. Recent 
UV and X-ray absorption line studies suggest that intermediate-temperature 
ionized gas is present in galaxy groups, with local densities perhaps as 
high as $10^{-4}$ cm$^{-3}$ (Fang et al. 2002; 
Nicastro et al. 2002; Sembach et al. 2003). The origin of such features 
is not fully established, but they are most likely due to a combination 
of cosmic structure growth and disturbances associated with galaxies 
(e.g., Oort 1970; Dav\'e et al. 2001).

\subsection{Ionized gas}\label{Sect_ion}

The absence of neutral gas does not exclude the presence of ionized gas
in dSphs, and so surveys also have been made for diffuse ionized 
gas in a few dSph.  An ultraviolet absorption line search for an
ionized corona with $\ge10^5$ K
around the dSph Leo\,I, which experienced star formation until $\sim 2$
Gyr ago, was unsuccessful (Bowen et al.\ 1997; column density limit
$< 2 \cdot 10^{17}$ cm$^{-2}$).  Similarly, 
a search for a $10^6$ K component in the dSph Fornax, where star formation
ceased 100 -- 200 Myr ago, yielded no detection
(Gizis, Mould, \& Djorgovski 1993).  

However, gas may be present at
lower ionization levels 
(associated with gas at T $\le 10^4$ K) and hence have escaped detection. 
Such gas would likely have a kinetic temperature comparable to the 
velocity dispersion of the dSphs and stay bound to the dSphs. 
Sources of ionization would include the distance-dependent
Galactic UV radiation field, which would work for dSphs
within a few hundred kpc distance from massive spiral galaxies, or 
recent star-formation events (Lin \& Murray 1998; 
Mashchenko, Carignan, \& Bouchard 2002).
Internal sources of ionization that could operate in
isolated dSphs, such as Tucana and Cetus, include SNe Ia heating (Burkert
\& Ruiz-Lapuente 1997), hot white
dwarfs (Dupree \& Raymond 1983), or UV-bright post-asymptotic giant
branch stars (de Boer 1985).

Ionized gas at $\sim 10^4$ K would not have
been seen in previous surveys, and even the 
most sensitive emission line observations can detect only 
relatively large amounts of diffusely distributed ionized gas in 
Galactic dSphs.  Such a gas component, if it exists and if
it accounts for a significant fraction of the missing ISM in dSphs,
would avoid the problem of finding gas {\it removal} mechanisms.  

One would then still be faced with the question of explaining why certain
dwarfs, the dIrrs and some dEs, contain neutral gas, while the ISM remains 
fully ionized in others with fewer OB stars or other obvious internal 
ionization sources.
However, if the ISM {\em were} to become ionized in dSphs,
then it would expand, have lower densities, and therefore would be more 
vulnerable to ram pressure
stripping. Such a process may have occurred during the re-ionization of 
the universe, potentially cutting off the supplies of gas for star formation 
in dwarf galaxies (see Thoul \& Weinberg 1996; Barkana \& Loeb 1999; 
Bullock et al.\ 2000). However, if this effect were too widespread, then 
it would not be consistent with the wide range of stellar population 
ages and gas content found in 
many Local Group dwarfs. We plan to return to a discussion of this issue 
in a later paper.

\subsection{Gas Removal Summary}

Evidently small galaxies might lose their gas early-on due to the 
effects of re-ionization, or they could be stripped of gas when they 
venture too close to a giant galaxy, particularly if it is in a starburst 
phase, as may have been common when galaxies were young. 
Near the present-day Local Group giants the combination 
of tides and ram pressure stripping, likely with the aid of leaking 
ionizing radiation (Mashchenko, Carignan, \& Bouchard 2002), can 
clean gas from small galaxies. We further expect that galaxies with relatively 
low masses and less ISM will be more readily relieved of their gas  
when in proximity to giant systems. 
Thus in the giant galaxy stripping model, dwarf galaxies that 
evolved into nearly gas-free states as they came near a giant 
galaxy were most readily converted into dSphs. This model predicts 
that {\it no} gas-free dSph should be found outside of galaxy groups, and 
none are known at the present time. The intermediate cases 
presented by Tucana and 
Cetus, dSphs far from any large galaxies, then are critical. 
They either may be runaways from the inner 
Local Group, or, if the IGM is clumpy, stripped by dense 
areas of the Local Group's IGM.

\section{Summary and Conclusions}\label{Sect_Conc}

When dSphs formed most of 
their stars, they must have contained substantial amounts
of interstellar gas, and would presumably have resembled 
present-day dIrrs or related classes of galaxies.  
Yet somehow star formation ceased and gas was lost and these 
objects evolved into dSphs. 
That this process took place at different times is clear from 
the range in mean stellar population ages in the nearby dSph 
galaxies. Continuity then suggests that the birth of dSphs should 
be an ongoing process, and that ``pre-dSph'' galaxies may exist 
in or near the Local Group.
This leads to the question of 
whether plausible dSph progenitors can
be identified among well-studied dwarf galaxies with 
resolved stellar populations in our 
immediate cosmic neighborhood, and whether evidence exists for 
a continuing transformation from gas-rich to gas-poor dwarfs.  An 
alternative scenario would be that dSphs form a separate class of their
own, with properties pre-defined at birth, as suggested
by Binggeli (1994) and supported by Skillman \& Bender (1995).

Our analysis of the luminosity-metallicity relation based on 
chemical abundances of red giants shows the well-known
separation between dIrrs and dSphs in the sense that dIrrs are too
metal-poor for their luminosity as compared to dSphs.  Mere fading
would not turn typical dIrrs into typical dSphs, 
since they would also fade in 
surface brightness, resulting in surface brightnesses below those
of dSphs at the same luminosity. Conversely, the rapid 
initial enrichment that old populations in dSphs experienced is not
observed in dIrrs, where star formation and enrichment proceed
slowly over a Hubble time.  The dIrrs also would need to
lose their angular momentum and interstellar medium, 
since neither rotation nor gas are detected in 
typical dSphs.  Hence transformation from dIrrs to dSphs 
through gas loss alone appears to
be unlikely, and in this regard our conclusions parallel those 
of a number of earlier studies (e.g., Binggeli 1986; Thuan 1986). 

The dIrr/dSph transition-type galaxies exhibit stellar and
optical structural properties
consistent with those of dSphs, while containing  H\,{\sc i} gas. 
The dIrr/dSph systems also fall near 
the dSph locus in the luminosity-metallicity relation.  Like dSphs,
they show very little or, more commonly, no rotation in their 
H\,{\sc i}, and have stellar 
populations whose properties closely resemble those in a number
of dSphs.  Their present-day star-formation rates are very low, and a
complete cessation of stellar births 
would not significantly alter their position in the 
luminosity-metallicity diagram. The transition-type dwarfs
appear to be gas-rich examples of dSph systems.  As none is in danger 
of complete gas exhaustion due to ongoing star formation, a gas-cleaning 
process seems necessary to convert them into dSphs. 
Evidently early
astration was comparatively rapid in both transition-type and dSph galaxies,
so their internal conditions near the time of their
formation were probably factors in defining their
subsequent evolution.
We therefore find the transition-type dwarfs to
constitute plausible progenitors of dSphs.  

We briefly considered various mechanisms whereby dwarfs could lose their
interstellar matter, and discussed possible examples of 
ongoing gas removal.  Intrinsic
processes currently seem to play a negligible role, 
and tidal effects alone appear to
be incapable of having completely removed the 
gas from dSphs (unless the 
dSph orbits, which are still unknown, are highly eccentric and we have 
missed the rare gas stripping episodes; see Mayer et al. 2001a,b).
The most plausible generic gas removal mechanism appears to be ram pressure 
stripping, and at present this will be effective in 
removing primordial gas reserves only when dwarfs pass  
near giant galaxies.  This in combination with tidal effects may 
explain the clustering of dSphs around the Local Group spirals. 
We discuss the case of the dSph Fornax, which lost its gas 
during the past 100 Myr and which is approaching its perigalacticon,
but find ram pressure stripping through a homogeneous IGM to be little
likely to be responsible for the removal of Fornax' ISM.

Beyond the halos of spirals, a smooth Local Group IGM, with 
its low estimated upper limit gas density of $n_H < 10^{-5}$~cm$^{-3}$,
will have little impact on the typical H\,{\sc i} disks of even 
small dwarfs (see Section \ref{Sect_ram}
and Mayer et al. 2001a). Yet we find a mixture 
of gas-deficient dSphs (Cetus and Tucana) along with gassy dIrr galaxies in 
the outer regions of the Local Group.  Due to gas return from 
dying stars, some process must act to have kept neutral gas from collecting in 
the outer Local Group dSphs, even if they were stripped of gas during 
a close passage to a spiral or some other past anomalous event. 
We find that a low density IGM can keep stellar 
mass loss from collecting. We also speculate that 
inhomogeneities in the Local Group IGM, such that some regions have densities 
of $n_H > 10^{-5}$~cm$^{-3}$, could strip dwarfs with low intrinsic gas 
densities. This option receives support from recent indications 
for inhomogeneities in the Local Group IGM derived from ultraviolet and 
X-ray absorption line measurements.  
The Local Group gas-stripping  
scenario implies that gas-free dwarf galaxies will not be found in locations 
without an IGM; isolated dSph galaxies outside of groups or clusters 
of galaxies should not exist.

Alternatively,
some of the missing gas in dSphs may simply be ionized 
and at temperatures near $10^4$
K, which would explain why it was not detected in  H\,{\sc i} 21-cm 
line surveys nor in UV absorption line studies.  A combination of 
background and internal stellar ionization
sources will be sufficient to sustain a relatively massive ionized 
interstellar medium, even in isolated, 
H\,{\sc i}-deficient dSphs, such as Tucana.  Sensitive Fabry-P\'erot techniques
may make the detection of ionized interstellar matter in dSphs possible,  
if such gas exists (Gallagher et al.\ 2003). 
Since ionization will reduce gas densities (and gravitational binding 
energies), it would also make it easier to strip interstellar matter 
from dwarfs. A combination of ionization and low-level 
ram pressure stripping should prevent the accumulation of 
gas lost from evolving stars in Local Group dSphs.

Environmental effects clearly have played a role in shaping the 
morphology--distance relation and the gas content of dwarf galaxies,
but are certainly not the sole determining factor.  For instance,
while the gas content of low-mass dwarfs increases with increasing
distance from a massive galaxy, transition-type dwarf galaxies with
their low H\,{\sc i} masses exist both in the wider surroundings of
massive galaxies as well as in the field.  
Furthermore, the current environment 
may have little connection with 
the formation of isolated dSphs like Tucana or Cetus. In this picture the 
Local Group dSph galaxies arose from a combination of factors. One 
parameter is closeness to one of the giant galaxies, but we argue that 
``genetics'' is also important. The metallicities in older stars in 
dSphs suggest relatively rapid initial and mid-life star formation 
depleted gas supplies, making them vulnerable to stripping.
Transition dwarf dIrr/dSphs then are examples of 
this class of galaxy that have retained their cool gas, while the more 
leisurely evolving dIrrs are distinguished by having more gas and angular 
momentum while being less chemically evolved.

\acknowledgments

This research has made use of the NASA/IPAC Extragalactic Database (NED),
which is operated by the Jet Propulsion Laboratory, California Institute 
of Technology, under contract with the National Aeronautics and Space 
Administration.  EKG would like to thank Klaas de Boer for his 
hospitality during her stay at the Sternwarte of the University of Bonn,
where part of this work was done.
JSG acknowledges support from the National Science Foundation through grant 
AST 98-03018 to the University of Wisconsin-Madison and from the University's  
Graduate School. We thank Hans-Walter Rix for the hospitality given 
to JSG at the MPIA while this work was in progress.

\onecolumn

\begin{deluxetable}{llccrrrccrr}
\rotate
\tabletypesize{\scriptsize}
\tablecaption{Properties of nearby dwarf galaxies. \label{tbl-1}}
\tablewidth{0pt}
\tablehead{
\colhead{Name} & \colhead{Type}   & \colhead{$\alpha$ (J2000)}   &
\colhead{$\delta$ (J2000)} &
\colhead{D$_{\odot}$}  & \colhead{D$_{\rm near}$} & \colhead{M$_V$} &
\colhead{$\mu_V$}     & \colhead{[Fe/H]
}  & \colhead{$M_{\rm HI}$}   & \colhead{$L_V$}\\
& & $^{\rm h}$ $^{\rm m}$ $^{\rm s}$ & $^{\circ}$ $'$ $''$ & 
\colhead{kpc} & \colhead{Mpc} &
\colhead{mag} & mag arcsec$^{-2}$ & dex & \colhead{10$^6$ M$_{\odot}$} 
& \colhead{10$^6$ L$_{\odot}$} 
}
\startdata
Sgr       &dSph,t,N?& 18 55 03 & $-30$ 28 42 &  28$\pm$\,\, 3 &               & $-15.0$&$25.4\pm0.2$& $-0.5\pm0.8$ &$<0.01$:\, & 80.1 \\
LMC       &IrIII-IV & 05 23 35 & $-69$ 45 22 &  50$\pm$\,\, 2 &               & $-18.5$&$20.7\pm0.1$& $-0.6\pm0.5$ &  500.0\,\,\,\,\,\,  &1995.0 \\
SMC       &IrIV/IV-V& 00 52 49 & $-72$ 49 43 &  63$\pm$\,\,10 &               & $-17.1$&$22.1\pm0.1$& $-1.2\pm0.4$ &  420.0\,\,\,\,\,\,  &550.0 \\
UMi       & dSph    & 15 09 10 & $+67$ 12 52 &  69$\pm$\,\, 4 &               & $ -8.9$&$25.5\pm0.5$& $-1.9\pm0.7$ &$<0.007$ & 0.3 \\
Dra       & dSph    & 17 20 12 & $+57$ 54 55 &  79$\pm$\,\, 4 &               & $ -9.4$&$24.4\pm0.5$& $-2.0\pm0.7$ &$<0.008$ & 0.47 \\
Sex       & dSph    & 10 13 03 & $-01$ 36 53 &  86$\pm$\,\, 6 &               & $ -9.5$&$26.2\pm0.5$& $-1.9\pm0.4$ &$<0.03$\,\,  & 0.5 \\
Scl       & dSph    & 01 00 09 & $-33$ 42 33 &  88$\pm$\,\, 4 &               & $ -9.8$&$23.7\pm0.4$& $-1.5\pm0.5$ &$>0.09$:\, & 2.2 \\
Car       & dSph    & 06 41 37 & $-50$ 57 58 &  94$\pm$\,\, 5 &               & $ -9.4$&$25.5\pm0.4$& $-1.8\pm0.3$ &$<0.001$& 0.4 \\
For       & dSph    & 02 39 59 & $-34$ 26 57 & 138$\pm$\,\, 8 &               & $-13.1$&$23.4\pm0.3$& $-1.2\pm0.5$ & $<0.7$\,\,\,\,\,\,  &  15.5\\
Leo\,II   & dSph    & 11 13 29 & $+22$ 09 17 & 205$\pm$\,\,12 &               & $-10.1$&$24.0\pm0.3$& $-1.6\pm0.5$ & $<0.03$\,\, &  0.6\\
Leo\,I    & dSph    & 10 -8 27 & $+12$ 18 27 & 270$\pm$\,\,10 &               & $-11.9$&$22.4\pm0.3$& $-1.4\pm0.5$ &$<0.009$ & 4.8 \\
Phe       &dIrr/dSph& 01 51 06 & $-44$ 26 41 & 405$\pm$\,\,15 &               & $ -9.8$& ---        & $-1.9\pm0.4$ & 0.17?   & 0.9  \\
NGC\,6822 &dIrIV-V  & 19 44 56 & $-14$ 52 11 & 500$\pm$\,\,20 &               & $-16.0$&$21.4\pm0.2$& $-1.2\pm0.4$ &  140.0\,\,\,\,\,\,  & 94.4 \\
\tableline
M32       & dE2,N   & 00 42 42 & $+40$ 51 55 & 770$\pm$\,\,40 & $\sim0.0_{\rm M31}$&$-16.5$&$11.5\pm0.5$& $-1.1\pm0.6$ & $<2.5$\,\,\,\,\,\,  &383.0   \\
NGC\,205  & dE5p,N  & 00 40 22 & $+41$ 41 07 & 830$\pm$\,\,35 & 0.06$_{\rm M31}$  & $-16.4$&$20.4\pm0.4$& $-0.5\pm0.5$ & 0.39\,\,\,    &366.0   \\
And\,I    & dSph    & 00 45 40 & $+38$ 02 28 & 790$\pm$\,\,30 & 0.05$_{\rm M31}$  & $-11.8$&$24.9\pm0.1$& $-1.4\pm0.2$ &$<0.096$ & 4.2 \\
And\,III  & dSph    & 00 35 34 & $+36$ 29 52 & 760$\pm$\,\,70 & 0.07$_{\rm M31}$  & $-10.2$&$25.3\pm0.1$& $-1.7\pm0.2$ & 0.09?   & 1.0 \\
NGC\,147  & dE5     & 00 33 12 & $+48$ 30 29 & 755$\pm$\,\,35 & 0.10$_{\rm M31}$  & $-15.1$&$21.6\pm0.2$& $-1.1\pm0.4$ & $<.005$ &131.0   \\
And\,V    & dSph    & 01 10 17 & $+47$ 37 41 & 810$\pm$\,\,45 & 0.12$_{\rm M31}$  & $ -9.1$&$24.8\pm0.2$& $-1.9\pm0.1$ &  undet. & 0.4 \\
And\,II   & dSph    & 01 16 30 & $+33$ 25 09 & 680$\pm$\,\,25 & 0.16$_{\rm M31}$  & $-11.8$&$24.8\pm0.1$& $-1.5\pm0.3$ &  undet. & 2.4 \\
NGC\,185  & dE3p    & 00 38 58 & $+48$ 20 12 & 620$\pm$\,\,25 & 0.17$_{\rm M31}$  & $-15.6$&$20.1\pm0.4$& $-0.8\pm0.4$ & 0.13\,\,\,    &125.0   \\
Cas\,dSph & dSph    & 23 26 31 & $+50$ 41 31 & 760$\pm$\,\,70 & 0.22$_{\rm M31}$  & $-12.0$&$23.5\pm0.1$& $-1.5\pm0.2$ &  undet. & 5.0  \\
IC\,10    &dIrIV/BCD& 00 20 17 & $+59$ 18 14 & 660$\pm$\,\,65 & 0.25$_{\rm M31}$  & $-16.0$&$20.4\pm0.4$& $-1.3\pm0.4$ &   98.0\,\,\,\,\,\,  &160.0   \\
And\,VI   & dSph    & 23 51 46 & $+24$ 34 57 & 775$\pm$\,\,35 & 0.27$_{\rm M31}$  & $-11.3$&$24.3\pm0.1$& $-1.7\pm0.2$ &  undet. & 2.6  \\
LGS\,3    &dIrr/dSph& 01 03 53 & $+21$ 53 05 & 620$\pm$\,\,20 & 0.28$_{\rm M31}$  & $ -9.8$&$24.7\pm0.2$& $-1.7\pm0.3$ & 8.0\,\,\,\,\,\,     &1.3  \\
PegDIG    &dIrr/dSph& 23 28 36 & $+14$ 44 35 & 760$\pm$100& 0.41$_{\rm M31}$  & $-12.9$& ---        & $-2.0\pm0.3$ &   3.4\,\,\,\,\,\,   & 12.0   \\
IC\,1613  & dIrrV   & 01 04 47 & $+02$ 07 02 & 715$\pm$\,\,35 & 0.50$_{\rm M31}$  & $-15.3$&$22.8\pm0.3$& $-1.4\pm0.3$ &   58.0\,\,\,\,\,\,  & 63.6 \\
\tableline
Cet       & dSph    & 00 26 11 & $-11$ 02 40 & 775$\pm$\,\,50 & 0.68$_{\rm M31}$  & $-10.1$&$25.1\pm0.1$& $-1.7\pm0.2$ &  undet. & 0.9 \\
Leo\,A    & dIrrV   & 09 59 26 & $-02$ 46 37 & 800$\pm$\,\,40 & 0.98$_{\rm bary}$ & $-11.7$& ---        & $-2.1\pm0.4$ &  7.6\,\,\,\,\,\,    & 4.1 \\
WLM       & dIrrIV-V& 00 01 58 & $-15$ 27 39 & 945$\pm$\,\,40 & 0.84$_{\rm M31}$  & $-14.4$&$20.4\pm0.1$& $-1.4\pm0.4$ &   63.0\,\,\,\,\,\,  & 50.2 \\
Tuc       & dSph    & 22 41 49 & $-64$ 25 12 & 870$\pm$\,\,60 & 1.11$_{\rm bary}$ & $ -9.6$&$25.1\pm0.1$& $-1.7\pm0.2$ &$<0.015$ & 0.6 \\
DDO\,210  &dIrr/dSph& 20 46 52 & $-12$ 50 53 & 950$\pm$\,\,50 & 0.96$_{\rm bary}$ & $-10.9$&$23.0\pm0.3$& $-1.9\pm0.3$ &  2.7\,\,\,\,\,\,    & 0.8 \\
SagDIG    & dIrrV   & 19 29 59 & $-17$ 40 41 &1060$\pm$100& 1.18$_{\rm bary}$ & $-12.0$&$23.9\pm0.2$& $-2.3\pm0.4$ &   8.6\,\,\,\,\,\,   & 6.9 \\
\tableline
NGC\,3109 & dIrrIV-V& 10 03 07 & $-26$ 09 32 &1360$\pm$100& 1.75$_{\rm bary}$ & $-15.7$&$23.6\pm0.2$& $-1.7\pm0.4$ &  820.0\,\,\,\,\,\,    &160.0   \\
Ant       &dIrr/dSph& 10 04 04 & $-27$ 19 55 &1330$\pm$100&$\sim.04_{\rm N3109}$&$-11.2$& ---      & $-1.9\pm0.2$ &  0.72\,\,\,   &  2.4 \\
Sex\,A    & dIrrV   & 10 11 01 & $-04$ 41 34 &1440$\pm$\, 70& 0.53$_{\rm N3109}$& $-14.6$&$23.5\pm0.3$& $-1.9\pm0.4$ &   54.0\,\,\,\,\,\,    & 55.7 \\
Sex\,B    & dIrrIV-V& 10 00 00 & $+05$ 19 56 &1320$\pm$140& 0.73$_{\rm N3109}$& $-14.2$& ---        & $-2.1\pm0.4$ &   44.0\,\,\,\,\,\,    & 40.7 \\
\tableline
IC\,5152  &dIrIV-V  & 22 02 42 & $-51$ 17 44 &1700$\pm$150& 1.83$_{\rm bary}$ & $-14.8$& ---        & $-1.4\pm0.5$ &   67.0\,\,\,\,\,\,    & 70.3 \\
KKR\,25   &dIrr/dSph& 16 13 48 & $+54$ 22 16 &1860$\pm$120& 1.79$_{\rm bary}$ & $-10.5$&$24.0\pm0.2$& $-2.1\pm0.3$ &   1.0\,\,\,\,\,\,   &1.2  \\
GR8       & dIrrV   & 12 58 40 & $+14$ 13 03 &2200$\pm$300& 2.50$_{\rm bary}$ & $-11.6$&$22.3\pm0.2$& $-2.0\pm0.4$ &  9.6\,\,\,\,\,\,    & 3.4 \\
\\
\\
\\
\\
\\
\\
\\
\\
 \enddata
\tablecomments{
{References for Table 1:
The galaxy type nomenclature of
van den Bergh (1994) was adopted, adding ``N'' (nucleated)
for Sgr, NGC\,205, and M32.  
References for heliocentric distances: Grebel (2000) except
for LGS\,3 (Miller et al.\ 2001), Leo\,A (Dolphin et al.\ 2002; 
Schulte-Ladbeck et al.\ 2002), SagDIG (Karachentsev, Aparicio, \&
Makarova 1999), the Sextans-Antlia group (van den
Bergh 1999b), IC\,5152 (Zijlstra \& Minniti 1999), KKR\,25 (Karachentsev 
et al.\ 2001), and GR\,8 (Dohm-Palmer et al.\ 1998).
References for absolute $V$ magnitudes: van den Bergh
(2000) except for Sgr (Majewski et al.\ 1998), Draco (Odenkirchen et al.\
2001), IC\,10 
(Richer et al.\ 2001), LGS\,3 
(Miller et al.\ 2001), Sextans-Antlia group (van den Bergh 1999b), 
IC\,5152 (Zijlstra \& Minniti 1999), KKR\,25 (Karachentsev et al.\
2001), and GR\,8 (Mateo 1998).  Central surface brightnesses: 
Mateo (1998) except for Draco (Aparicio et al.\
2001), 
And\,V, VI, and Cas\,dSph (Caldwell 1999), 
IC\,10 (Richer et al.\ 2001), Cetus (Whiting, Hau, \& Irwin 1999), DDO\,210
(Lee et al.\ 1999), KKR\,25 (Karachentsev et al.\ 2001).  
Red giant branch metallicities:
Grebel (2000) except for Sgr (Cole 2001), Leo\,I and II 
(Smecker-Hane, Bosler, \& Stetson 2003), Leo\,A (Schulte-Ladbeck et al.\
2002), UMi, Dra, Sex (Shetrone et al.\ 2001), Scl, For (Tolstoy et al.\
2003), KKR\,25 (Karachentsev et al.\ 2001), and our ongoing 
re-determination of red giant branch metallicities using archival
WFPC2 images (Harbeck et al.\ 2001).  
H\,{\sc i} masses were recalculated from the integrated flux density
or from upper limits assuming a uniform H\,{\sc i} column density within 
the tidal radius (Section \protect{\ref{Sect_baryon}}) using the updated
heliocentric distances given above.   
References: Sgr: Koribalski, Johnston, \& Otrupcek (1994,
assuming an angular size of $10\degr \times 4\degr$), LMC: Kim et al.\
(1998), SMC: Stanimirovi\'c et al.\ (1999), UMi, Dra, Sex, Leo\,I: Young
(2000), For, Leo\,II: Young (1999), Car: Mould et al.\ (1994), Scl, 
And\, II, And\,III, And\,V, And\,VI, Cas\,dSph, DDO\,210, LGS\,3, 
PegDIG: Blitz \& Robishaw (2000), And\,I: Thuan \& Martin (1979),
Cet: Huchtmeier, Karachentsev, \& Karachentseva (2001),  
Phe: St-Germain et al.\ (1999), Tuc: Oosterloo et al.\ (1996), Ant:
Barnes \& de Blok (2001), KKR\,25: Karachentsev et al.\ (2001), and
Mateo (1998) for the remainder. 
}
}
\end{deluxetable}

\clearpage

%\clearpage

%\clearpage

%\hfill

\clearpage


\begin{thebibliography}{}

\bibitem[Andersen \& Burkert(2000)]{And2000} Andersen, R.-P.~\& Burkert, 
        A.\ 2000, \apj, 531, 296 

\bibitem[Aparicio, Carrera, \& Mart\'{\i}nez-Delgado(2001)]{Apa2001} 
        Aparicio, A., Carrera, R., \& Mart\'{\i}nez-Delgado, D.\ 2001, 
        \aj, 122, 2524

\bibitem[Barnes \& de Blok(2001)]{Bar2001} Barnes, D.~G.~\& de Blok, 
        W.~J.~G.\ 2001, \aj, 122, 825

\bibitem[Barkana \& Loeb(1999)]{Bark1999} Barkana, R.~\& Loeb, A.\ 1999, 
        \apj, 523, 54 

\bibitem[Bersier \& Wood(2002)]{Ber2002} Bersier, D.~\& Wood, 
        P.~R.\ 2002, \aj, 123, 840 

\bibitem[Binggeli(1986)]{Binggeli1986} Binggeli, B.\ 1986, in Star-Forming
        Dwarf Galaxies and Related Objects, eds.\ D. Kunth, T.~X. Thuan, \& 
        J.~T.~T. Thuan (Gif-sur-Yvette: Editions Fronti\`eres), 52

\bibitem[Binggeli(1994)]{Binggeli1994} Binggeli, B. 1994, in Panchromatic 
        View of Galaxies-their Evolutionary Puzzle, eds. G. Hensler, Ch. 
        Theis, J. S. Gallagher (Gif-sur-Yvette: Editions Fronti\'eres), 173

\bibitem[Blitz \& Robishaw(2000)]{Blitz2000} Blitz, L.~\& Robishaw, T.\ 
        2000, \apj, 541, 675

\bibitem[Bonifacio et al.(2000)]{Boni2000} Bonifacio, P., Hill, V., Molaro, 
        P., Pasquini, L., Di Marcantonio, P., \& Santin, P.\ 2000, 
        \aap, 359, 663 

\bibitem[Bowen et al.(1997)]{Bow1997} Bowen, D.~V., Tolstoy, E., Ferrara, 
        A., Blades, J.~C., \& Brinks, E.\ 1997, \apj, 478, 530

\bibitem[Braun \& Burton(2001]{Bra2001} Braun, R.~\& Burton, 
        W.~B.\ 2001, \aap, 375, 219 

\bibitem[Bullock, Kravtsov, \& Weinberg(2000)]{Bull2000} Bullock, J.~S., 
        Kravtsov, A.~V., \& Weinberg, D.~H.\ 2000, \apj, 539, 517

\bibitem[Buonanno et al.(1998)]{Buon1998} Buonanno, R., Corsi, C.~E., 
        Zinn, R., Fusi Pecci, F., Hardy, E., \& Suntzeff, N.~B.\ 1998, 
        \apjl, 501, L33

\bibitem[Bureau \& Carignan (2002)]{Bure2002} Bureau, M. \& Carignan, C. 
	2002, \aj, 123, 1316

\bibitem[Burkert \& Ruiz-Lapuente (1997)]{Burk97} Burkert, A. \& 
	Ruiz-Lapuente 1997, \apj, 480, 297

\bibitem[Caldwell(1999)]{Cald1999} Caldwell, N.\ 1999, \aj, 118, 1230

\bibitem[Caldwell et al.(1998)]{Cald1998} Caldwell, N., Armandroff, T.~E., 
        Da Costa, G.~S., \& Seitzer, P.\ 1998, \aj, 115, 535 

\bibitem[Carigi, Hernandez, \& Gilmore(2002)]{Carigi2002} Carigi, L., 
        Hernandez, X. \& Gilmore, G. 2002, MNRAS, 334, 117

\bibitem[Carignan(1999)]{Car1999} Carignan, C.\ 1999, PASA, 16, 18 

\bibitem[Carignan et al.(1998)]{Car1998} Carignan, C., Beaulieu, S., 
        C{\^ o}t{\' e}, S., Demers, S., \& Mateo, M.\ 1998, \aj, 116, 1690

\bibitem[Carpenter(2000)]{Carp2000} Carpenter, J.~M.\ 2000, \aj, 120, 3139

\bibitem[Carraro et al. (2001)]{Carra2001} Carraro, G., Chiosi, C., Girardi, 
	L., \& Lia, C. 2001, \mnras, 327, 69

\bibitem[Charlot, Worthey, \& Bressan(1996)]{Char1996} Charlot, 
        S., Worthey, G., \& Bressan, A.\ 1996, \apj, 457, 625 

\bibitem[Chiu, Gnedin, \& Ostriker (2001)]{Chiu2001} Chiu, W. A., Gnedin, 
	N. Y., \& Ostriker, J. P. 2001, \apj, 563, 21

\bibitem[Cole(2001)]{Cole2001} Cole, A.~A.\ 2001, \apjl, 559, L17

\bibitem[C\^ot\'e, Oke, \& Cohen(1999)]{Cot1999} C\^ot\'e, P., Oke, J.~B., 
        \& Cohen, J.~G.\ 1999, \aj, 118, 1645

\bibitem[Courteau \& van den Bergh(1999)]{Cour1999} Courteau, S. \& van den 
        Bergh, S.\ 1999, \aj, 118, 337

\bibitem[Da Costa(1998)]{DaC1998}  Da Costa, G.~S. 1998, in Stellar
        astrophysics for the Local Group, VIII Canary Islands Winter
        School of Astrophysics, eds.\ A. Aparicio, A. Herrero, \& F. Sanchez
        (Cambridge: CUP), 351

\bibitem[Da Costa \& Armandroff(1990)]{DaC1990} Da Costa, G.~S.~\& 
        Armandroff, T.~E.\ 1990, \aj, 100, 162

\bibitem[Dav{\' e} et al.(2001)]{Dave2001} Dav{\' e}, R.~et al.\ 2001, 
        \apj, 552, 473 

\bibitem[Davies \& Phillipps(1988)]{Davies1988}
        Davies, J.~I., \& Phillipps, S. 1988, MNRAS, 233, 553

\bibitem[de Blok \& Walter(2000)]{deBW2000} de Blok, W.J.G., \& Walter, F. 
        2000, ApJ, 537, L95

\bibitem[de Blok et al.(2002)]{de Blok2002} de Blok, W.~J.~G., Zwaan, 
        M.~A., Dijkstra, M., Briggs, F.~H., \& Freeman, K.~C.\ 2002, \aap, 
        382, 43 

\bibitem[de Boer(1985)]{deB1985} de Boer, K.~S. 1985, A\&A, 142, 321

\bibitem[Dekel \& Silk 1986]{DS86} Dekel, A. \& Silk, J 1986, ApJ, 303, 39

\bibitem[Dohm-Palmer et al.(1998)]{Dohm1998} Dohm-Palmer, R.~C.~et al.\ 
        1998, \aj, 116, 1227

\bibitem[Dolphin et al.(2002)]{Dol2002} Dolphin, A.~E.~et al.\ 2002, \aj, 
        123, 3154 

\bibitem[Dupree \& Raymond(1983)]{Dup1983} Dupree, A.~K.~\& 
        Raymond, J.~C.\ 1983, \apjl, 275, L71 

\bibitem[Elmegreen \& Hunter(2000)]{Elm2000} Elmegreen, B.~G.~\& Hunter, 
        D.~A.\ 2000, \apj, 540, 814

\bibitem[Eskridge(1988)]{Esk1988} Eskridge, P.~B.\ 1988, ApL\&C, 26, 315 

\bibitem[Fang et al.(2002)]{Fang2002} Fang, T., Marshall, H.~L., 
        Lee, J.~C., Davis, D.~S., \& Canizares, C.~R.\ 2002, \apjl, 572, L127 

\bibitem[Ferguson \& Binggeli(1994)]{Ferg1994} Ferguson, H.~C.~\& 
        Binggeli, B.\ 1994, \aapr, 6, 67

\bibitem[Ferrara \& Tolstoy(2000)]{Ferr2000} Ferrara, A.~\& Tolstoy, E.\ 2000, 
        \mnras, 313, 291

\bibitem[Gallagher (1998)]{Gall98}Gallagher, J.~S. 1998, in The Magellanic
        Clouds and Other Dwarf Galaxies, eds. T. Richtler  \& J.~M. Braun
        (Aachen: Shaker Verlag), 25

\bibitem[Gallagher \& Wyse(1994)]{Gallagh1994} Gallagher, J.~S.~\& 
        Wyse, R.~F.~G.\ 1994, \pasp, 106, 1225

\bibitem[Gallagher et al.(1998)]{Gallagh1998} Gallagher, J.~S., Tolstoy, E., 
        Dohm-Palmer, R.~C., Skillman, E.~D., Cole, A.~A., Hoessel, J.~G., 
        Saha, A., \& Mateo, M.\ 1998, \aj, 115, 1869

\bibitem[Gallagher et al.(2003)]{Gallagh2003} Gallagher, J.S., Madsen, G., 
        Reynolds, R., Grebel, E.K., \& Smecker-Hane, T.
        ApJ Lett, submitted

\bibitem[Gallart et al.(2001)]{Gall2001} Gallart, C., Mart\'{\i}nez-Delgado, 
        D., G{\' o}mez-Flechoso, M.~A., \& Mateo, M.\ 2001, \aj, 121, 2572

\bibitem[Gerola, Carnevali, \& Salpeter(1983)]{Ger1983} Gerola, 
        H., Carnevali, P., \& Salpeter, E.~E.\ 1983, \apjl, 268, L75 

\bibitem[Gizis, Mould, \& Djorgovski(1993)]{Giz93} Gizis, J.~E., Mould, J.~R., 
        \& Djorgovski, S.\ 1993, \pasp, 105, 871


\bibitem[Grebel(1997)]{Greb1997} Grebel, E.~K.\ 1997, Reviews of Modern 
        Astronomy, 10, 29

\bibitem[Grebel(1999)]{Greb1999} Grebel, E.~K.\ 1999, in The Stellar Content
        of the Local Group, IAU Symp.\ 192, eds.\ P.\ Whitelock \& R.\
        Cannon (Provo: ASP), 17

\bibitem[Grebel(2000)]{Greb2000} Grebel, E.~K.\ 2000, in Star Formation from
        the Small to the Large Scale, 33rd ESLAB Symp., SP-445, eds.\ F.\ 
        Favata, A.~A.\ Kaas, \& A.\ Wilson (Noordwijk: ESA), 87

\bibitem[Grebel(2002)]{Greb2002} Grebel, E.~K.\ 2002, in Gaseous Matter in
        Galaxies and Intergalactic Space, 17th IAP Colloquium, ed.\ R.\
        Ferlet (Paris: Frontier Group), 171 

\bibitem[Grebel \& Guhathakurta(1999)]{GreGuh99} Grebel, E.~K. \& 
        Guhathakurta, P. 1999, ApJ, 511, L101

\bibitem[Grebel \& Stetson(1999)]{GreSte99}  Grebel, E.~K. \& Stetson, 
        P.~B. 1999, in The Stellar Content
        of the Local Group, IAU Symp.\ 192, eds.\ P.\ Whitelock \& R.\
        Cannon (Provo: ASP), 165

\bibitem[Gunn \& Gott (1972)]{GunGo72} Gunn, J. E. \& Gott, J. R., III 
        1972, \apj, 176, 1

\bibitem[Guhathakurta, Reitzel, \& Grebel(2000)]{Guh2000} Guhathakurta, P., 
        Reitzel, D.~B., \& Grebel, E.~K.\ 2000, \procspie, 4005, ed.\
        J.\ Bergeron, 168

\bibitem[Harbeck et al.(2001)]{Har2001} Harbeck, D.~et al.\ 
        2001, \aj, 122, 3092

\bibitem[Hirashita(1999)]{Hira1999} Hirashita, H.\ 1999, \aap, 344, L87 

\bibitem[Hirashita, Kamaya, \& Mineshige(1997)]{Hira1997} Hirashita, H., 
        Kamaya, H., \& Mineshige, S.\ 1997, \mnras, 290, L33 

\bibitem[Holtzman, Smith, \& Grillmair(2000)]{Holtz2000} Holtzman, J.~A., 
        Smith, G.~H., \& Grillmair, C.\ 2000, \aj, 120, 3060

\bibitem[Huchtmeier, Karachentsev, \& Karachentseva(2001)]{Huc2001} 
        Huchtmeier, W.~K., Karachentsev, I.~D., \& Karachentseva, V.~E.\ 
        2001, \aap, 377, 801

\bibitem[Hurley-Keller, Mateo, \& Nemec(1998)]{HKel1998} 
        Hurley-Keller, D., Mateo, M., \& Nemec J., 1998, \aj, 115, 1840

\bibitem[Hunter (1997)]{Hunt97} Hunter, D. A. 1997, PASP, 109, 937

\bibitem[Hunter \& Gallagher(1985)]{Hunt1985} Hunter, D.~A.~\& Gallagher, 
        J.~S.\ 1985, \apjs, 58, 533

\bibitem[Ikuta \& Arimoto(2002)]{Iku02} Ikuta, C. \& Arimoto, N. 2002, 
        A\&A, 391, 55

\bibitem[Irwin \& Tolstoy(2002)]{Irw02} Irwin, M., \& Tolstoy, E. 2002,
        MNRAS, 336, 643

\bibitem[Jungwiert, Combes, \& Palou{\v s}(2001)]{Jung2001} 
        Jungwiert, B., Combes, F., \& Palou{\v s}, J.\ 2001, \aap, 376, 85 

\bibitem[Karachentsev, Aparicio, \& Makarova(1999)]{Kar1999} Karachentsev, 
        I., Aparicio, A., \& Makarova, L.\ 1999, \aap, 352, 363

\bibitem[Karachentsev et al.(2001)]{Kar2001} Karachentsev, I.~D.~et al.\ 
        2001, \aap, 379, 407

\bibitem[Kim et al.(1998)]{Kim1998} Kim, S., Staveley-Smith, L., Dopita, 
        M.~A., Freeman, K.~C., Sault, R.~J., Kesteven, M.~J., \& McConnell, 
        D.\ 1998, \apj, 503, 674

\bibitem[Kleyna et al.(2002)]{Kleyna2002} Kleyna, J., Wilkinson, 
        M.~I., Evans, N.~W., Gilmore, G., \& Frayn, C.\ 2002, \mnras, 330, 792

\bibitem[Klypin et al.(1999)]{Klyp1999} Klypin, A., Kravtsov, A.~V., 
        Valenzuela, O., \& Prada, F.\ 1999, \apj, 522, 82 

\bibitem[Knapp, Kerr, \& Bowers(1978)]{Kna1978} Knapp, G.~R., Kerr, F.~J., 
        \& Bowers, P.~F.\ 1978, \aj, 83, 360

\bibitem[Knapp, Kerr, \& Williams(1978)]{KKW1978} Knapp, G.~R., Kerr, F.~J., 
        \& Williams, B.~A.\ 1978, \apj, 222, 800

\bibitem[Koribalski, Johnston, \& Otrupcek(1994)]{Kor1994} Koribalski, B., 
        Johnston, S., \& Otrupcek, R.\ 1994, \mnras, 270, L43

\bibitem[Kormendy(1985)]{Kor1985} Kormendy, J.\ 1985, \apj, 295, 73

\bibitem[Lee, Freedman, \& Madore(1993)]{Lee1993} Lee, M.~G., Freedman, 
        W.~L., \& Madore, B.~F.\ 1993, \apj, 417, 553

\bibitem[Lee et al.(1999)]{Lee1999} Lee, M.~G., Aparicio, A., Tikhonov, N., 
        Byun, Y., \& Kim, E.\ 1999, \aj, 118, 853

\bibitem[Lin \& Faber(1983)]{Lin1983} Lin, D.~N.~C.~\& Faber, 
        S.~M.\ 1983, \apjl, 266, L21

\bibitem[Lin \& Murray(1998)]{Lin1998} Lin, D.~N.~C.\ \& Murray, S.~D.\ 
        1998, in Dwarf Galaxies and Cosmology, XXXIIIrd Rencontres de
        Moriond, eds.\ T.~X.Thuan, C.\ Balkowski, V.\ Cayatte, \&
        J.~T.~T. Van (Paris: Editions Fronti\'eres), 433

\bibitem[Lo, Sargent, \& Young(1993)]{Lo1993} Lo, K.~Y., Sargent, W.~L.~W., 
        \& Young, K.\ 1993, \aj, 106, 507

\bibitem[Mac Low \& Ferrara(1999)]{Mac1999} Mac Low, M.~\& Ferrara, A.\ 
        1999, \apj, 513, 142 

\bibitem[Majewski et al.(2000)]{Maj2000} Majewski, S.~R., Ostheimer, J.~C., 
        Patterson, R.~J., Kunkel, W.~E., Johnston, K.~V., \& Geisler, 
        D.\ 2000, \aj, 119, 760


\bibitem[Mart\'{\i}nez-Delgado et al.(2001)]{Mart2001} Mart\'{\i}nez-Delgado, 
        D., Alonso-Garc\'{\i}a, J., Aparicio, A., \& G{\' o}mez-Flechoso, 
        M.~A.\ 2001, \apjl, 549, L63 

\bibitem[Mashchenko, Carignan, \& Bouchard(2002)]{Mash2002} Mashchenko, S., 
        Carignan, C., \& Bouchard, A.\ 2002, ApJ, submitted 
        (astro-ph/0203317)

\bibitem[Mateo(1998)]{Mat1998} Mateo, M.~L.\ 1998, \araa, 36, 435

\bibitem[Matzner \& McKee(2000)]{Matz2000} Matzner, C.~D.~\& McKee, C.~F.\ 
        2000, \apj, 545, 364

\bibitem[Matthews, van Driel, \& Gallagher(1998)]{Matt1998} Matthews, L.~D., 
        van Driel, W., \& Gallagher, J.~S.\ 1998, \aj, 116, 2196 

\bibitem[Mayer et al.(2001a)]{May2001a} Mayer, L., Governato, F., Colpi, M., 
        Moore, B., Quinn, T., Wadsley, J., Stadel, J., \& Lake, G.\ 
        2001a, \apj, 559, 754  

\bibitem[Mayer et al.(2001b)]{May2001b} Mayer, L., Governato, 
        F., Colpi, M., Moore, B., Quinn, T., Wadsley, J., Stadel, J., 
        \& Lake, G.\ 2001b, \apjl, 547, L123

\bibitem[McKay et al.(2002)]{McKay2002} McKay, T.~A.~et al.\ 
        2002, \apjl, 571, L85 

\bibitem[Miller et al.(2001)]{Mil2001} Miller, B.~W., Dolphin, A.~E., 
        Lee, M.~G., Kim, S.~C., \& Hodge, P.\ 2001, \apj, 562, 713

\bibitem[Minniti \& Zijlstra(1996)]{Min1996} Minniti, D., \& Zijlstra,
        A.~A. 1996, ApJ, 467, L13

\bibitem[Moore et al. (1999)]{Moor1999} Moore, B, Ghigna, S., Governato, 
	F., Lake, G., Quinn, T., Stadel, J., \& Tozzi, P. 1999, \apj, 
	524, L19

\bibitem[Mori \& Burkert(2000)]{Mori2000} Mori, M.~\& Burkert, A.\ 2000, 
        \apj, 538, 559 

\bibitem[Mould et al.(1990)]{Mou1990} Mould, J.~R., Bothun, G.~D., Hall, 
        P.~J., Staveley-Smith, L., \& Wright, A.~E.\ 1990, \apjl, 362, L55

\bibitem[Murali(2000)]{Mur2000} Murali, C.\ 2000, \apjl, 529, L81 

\bibitem[Nicastro et al.(2002)]{Nica2002} Nicastro, F.~et al.\ 2002, \apj, 
        573, 157 

\bibitem[Odenkirchen et al.(2001)]{Oden2001} Odenkirchen, M.~et al.\ 2001, 
        \aj, 122, 2538

\bibitem[Oort(1970)]{Oort1970} Oort, J.~H.\ 1970, \aap, 7, 381 

\bibitem[Oosterloo, Da Costa, \& Staveley-Smith(1996)]{Oost1996} Oosterloo, 
        T., Da Costa, G.~S., \& Staveley-Smith, L.\ 1996, \aj, 112, 1969

\bibitem[Pedraz et al. (2002)]{Ped02} Pedraz, S., Gorgas, J., Cardiel, N. 
        S\'anchez-Bl\'azquez, P., \& Guzm\'an, R. 2002, MNRAS, 322, L59

\bibitem[Phillipps, Edmunds, \& Davies (1990)]{Phil90} Phillipps, S., 
        Edmunds, M. G., \& Davies, J. I. 1990, \mnras, 244, 168


\bibitem[Piatek et al.(2002)]{Piatek2002} Piatek, S.~et al.\ 2002, 
        \aj, 124, 3198 

\bibitem[Putman et al.(2002)]{Put2002} Putman, M.~E.~et al.\ 
        2002, \aj, 123, 873 

\bibitem[Quilis \& Moore(2001)]{Quil2001} Quilis, V.~\& Moore, B.\ 2001, 
        \apjl, 555, L95 

\bibitem[Recchi, Matteucci, \& S'Ercole (2001)]{Recchi2001} Recchi, S., 
	Matteucci, F., \& D'Ercole, A. 2001, \mnras, 322, 800

\bibitem[Richer, McCall, \& Arimoto(1997)]{Rich1997} Richer, M.~G., McCall, 
        M.~L., \& Arimoto, N.\ 1997, \aaps, 122, 215

\bibitem[Richer, McCall, \& Stasinska(1998)]{Rich1998} Richer, M., McCall, 
        M.~L., \& Stasinska, G.\ 1998, \aap, 340, 67

\bibitem[Richer et al.(2001)]{Rich2001} Richer, M.~G.~et al.\ 2001, \aap, 
        370, 34

\bibitem[Robishaw, Simon, \& Blitz(2002)]{Robi2002} Robishaw, T., Simon,
        J.~D., \& Blitz, L. 2002, \apj, 580, L129

\bibitem[Sarajedini et al.(2002)]{Sara2002} Sarajedini, A.~et 
        al.\ 2002, \apj, 567, 915 

\bibitem[Scannapieco, Ferrara, \& Broadhurst (2000)]{Sca2000} Scannapieco, E., 
        Ferrara, A., \& Broadhurst, T. 2000, \apj, 536, L11

\bibitem[Schulte-Ladbeck et al.(2002)]{Schu2002} Schulte-Ladbeck, R.~E., 
        Hopp, U., Drozdovsky, I.~O., Greggio, L., \& Crone, M.~M.\ 2002, 
        AJ, 124, 896

\bibitem[Schulz et al.(2002)]{Schulz2002} Schulz, J., Alvensleben, U.~F., 
        M{\"o}ller, C.~S., \& Fricke, K.~J.\ 2002, \aap, 392, 1 

\bibitem[Sembach et al.(2003)]{Sem2003} Sembach, K.~R.~et al.\ 2003, 
        ApJS, submitted (astro-ph/0207562)

\bibitem[Shetrone, C\^ot\'e, \& Sargent(2001)]{Shet2001} Shetrone, M.~D., 
        C\^ot\'e, P., \& Sargent, W.~L.~W.\ 2001, \apj, 548, 592

\bibitem[Shetrone et al.(2003)]{Shet2003} Shetrone, M., Venn, 
        K.~A., Tolstoy, E., Primas, F., Hill, V., \& Kaufer, A.\ 2003,
        AJ, in press (astro-ph/0211167)

\bibitem[Silk, Wyse, \& Shields(1987)]{Silk1987} Silk, J., Wyse, R.~F.~G., 
        \& Shields, G.~A.\ 1987, \apjl, 322, L59

\bibitem[Skillman \& Bender(1995)]{Skill1995} Skillman, E.~D.~\& Bender, R.\ 
        1995, RMxAC, 3, 25 

\bibitem[Smecker-Hane (1997)]{Sme1997} Smecker-Hane, T. 1997, in Star Formation 
        Near and far, Seventh Astrophysics Conference, AIP Conf. proc. 393, 
        eds.\ S.\ Holt \& L.\ G.\ Mundy (New York: AIP), 571

\bibitem[Smecker-Hane, Bosler, \& Stetson(2003)]{Sme2003} 
        Smecker-Hane, T. A., Bosler, T. L., \& Stetson, P. B. 2003, preprint

\bibitem[Smecker-Hane \& mCWilliam(2003)]{SmeMc2003}
        Smecker-Hane, T. A. \& McWilliam, A. 2003, preprint 

\bibitem[Sofue(1994)]{Sofue1994} Sofue, Y. 1994, ApJ, 423, 207

\bibitem[Stanimirovi\'c et al.(1999)]{Stan1999} Stanimirovi\'c, S., 
        Staveley-Smith, L., Dickey, J.~M., Sault, R.~J., \& Snowden, S.~L.\ 
        1999, \mnras, 302, 417

\bibitem[Stanimirovi\'c et al.(2002)]{Stan2002} Stanimirovi\'c, S., 
        Dickey, J.~M., Kr{\v c}o, M., \& Brooks, A.~M.\ 2002, \apj, 576, 773 

\bibitem[Stetson, Hesser, \& Smecker-Hane(1998)]{Stet1998PASP} Stetson, P.~B., 
        Hesser, J.~E., \& Smecker-Hane, T.~A.\ 1998, \pasp, 110, 533

\bibitem[St-Germain et al.(1999)]{StG1999} 
        St-Germain, J., Carignan, C., C{\^ o}te, S., \& Oosterloo, T.\ 
        1999, \aj, 118, 1235

\bibitem[Suntzeff et al.(1993)]{Sun1993} Suntzeff, N.~B., Mateo, M., 
        Terndrup, D.~M., Olszewski, E.~W., Geisler, D., \& Weller, W.\ 1993, 
        \apj, 418, 208

\bibitem[Tamura \& Hirashita (1999)]{Tam1999} Tamura, N. \& Hirashita, H. 
	1999, \apj, 525, L17

\bibitem[Thoul \& Weinberg(1996)]{Thoul1996} Thoul, A.~A.~\& Weinberg, 
        D.~H. 1996, \apj, 465, 608 

\bibitem[Thuan(1986)]{Thuan1986} Thuan, T.~X.\ 1986, in Star-Forming
        Dwarf Galaxies and Related Objects, eds.\ D. Kunth, T.~X. Thuan, \&
        J.~T.~T. Thuan (Gif-sur-Yvette: Editions Fronti\`eres), 105

\bibitem[Thuan \& Martin(1979)]{Thu1979} Thuan, T.~X.~\& Martin, G.~E.\ 
        1979, \apjl, 232, L11

\bibitem[Tolstoy et al.(1998)]{Tol1998} Tolstoy, E., Gallagher, J., 
        Cole, A. A., Hoessel, J. G., Saha, A. Dohm-Palmer, R. C., 
        Skillman, E. D., Mateo, M., \& Hurley-Keller, D. 1998, AJ, 116, 1244

\bibitem[Tolstoy et al.(2001)]{Tol2001} Tolstoy, E., Irwin, M.~J., Cole, 
        A.~A., Pasquini, L., Gilmozzi, R., \& Gallagher, J.~S.\ 2001, 
        \mnras, 327, 918

\bibitem[Tolstoy et al.(2003)]{Tol2003} Tolstoy, E., Venn, K.~A., 
        Shetrone, M., Primas, F., Hill, V., Kaufer, A., \& Szeifert, T.\ 
        2003, AJ, in press (astro-ph/0211168)

\bibitem[van den Bergh(1994)]{vdB1994} van den Bergh, S.\ 1994, \apj, 428, 617

\bibitem[van den Bergh(1999a)]{vdB1999a} van den Bergh, S.\ 1999a, A\&ARv, 9,
        273

\bibitem[van den Bergh(1999b)]{vdB1999b} van den Bergh, S.\ 1999b, \apjl, 
        517, L97

\bibitem[van den Bergh(2000)]{vdB2000} van den Bergh, S.\ 2000, The Galaxies
        of the Local Group, Cambridge Astrophysics Series Vol.\ 35
        (Cambridge: CUP)

\bibitem[van Driel et al.(2002)]{Driel2002} van Driel, W.~et al.\ 
        2002, A\&A, in press (astro-ph/0211181)

\bibitem[van Zee et al.(1997)]{Zee1997} van Zee, L., Haynes, M.~P., Salzer, 
        J.~J., \& Broeils, A.~H.\ 1997, \aj, 113, 1618

\bibitem[Wakker(2001)]{Wak2001} Wakker, B.P. 2001, ApJS, 136, 436

\bibitem[Whiting, Hau, \& Irwin(1999)]{Whit1999} Whiting, A.~B., 
        Hau, G.~K.~T., \& Irwin, M.\ 1999, \aj, 118, 2767

\bibitem[Young \& Lo(1997)]{Young1997} Young, L.~M.~\& Lo, K.~Y.\ 1997, 
        \apj, 490, 710

\bibitem[Young(1999)]{Young1999} Young, L.~M.\ 1999, \aj, 117, 1758

\bibitem[Young(2000)]{Young2000} Young, L.~M.\ 2000, \aj, 119, 188

\bibitem[Zijlstra \& Minniti(1999)]{Zij1999} Zijlstra, A.~A.~\& Minniti, D.\ 
        1999, \aj, 117, 1743

\bibitem[Zinn \& West(1984)]{Zinn1984} Zinn, R.~\& West, M.~J.\ 1984, 
        \apjs, 55, 45

\bibitem[Zwaan(2001)]{Zwaan2001} Zwaan, M.~A.\ 2001, MNRAS, 325, 1142


\end{thebibliography}
\end{document}